\documentclass[11pt]{amsart}
\reversemarginpar
\usepackage{amsmath} 
\usepackage{amssymb}
\usepackage{euscript}

\newcommand{\nwc}{\newcommand}

\nwc{\nwt}{\newtheorem}

\nwt{prop}{Proposition}
\nwt{lem}{Lemma}
\nwt{thm}{Theorem}
\nwt{cor}{Corollary}
\nwt{rem}{Remark}
\nwt{defin}{Definition} 

\nwc{\barr}{\begin{array}}
\nwc{\bal}{\begin{align}}
\nwc{\bequ}{\begin{equation}}
\nwc{\ben}{\begin{equation*}}
\nwc{\bea}{\begin{eqnarray}}
\nwc{\beq}{\begin{eqnarray}}
\nwc{\bean}{\begin{eqnarray*}}
\nwc{\beqn}{\begin{eqnarray*}}
\nwc{\beqast}{\begin{eqnarray*}}

\nwc{\earr}{\end{array}}
\nwc{\eal}{\end{align}}
\nwc{\eequ}{\end{equation}}
\nwc{\een}{\end{equation*}}
\nwc{\eea}{\end{eqnarray}}
\nwc{\eeq}{\end{eqnarray}}
\nwc{\eean}{\end{eqnarray*}}
\nwc{\eeqn}{\end{eqnarray*}}
\nwc{\eeqast}{\end{eqnarray*}}

\nwc{\bldm}{\boldmath} 
\nwc{\ubm}{\unboldmath} 
\nwc{\mf}{\mathbf} 
\nwc{\blds}{\boldsymbol} 
\nwc{\ml}{\mathcal} 

\nwc{\al}{\alpha}
\nwc{\vep}{\varepsilon}
\nwc{\ep}{\epsilon}
\nwc{\veps}{\varepsilon}
\nwc{\eps}{\epsilon}
\nwc{\vrho}{\varrho}
\nwc{\orho}{\bar\varrho}
\nwc{\vpsi}{\varpsi}
\nwc{\lamb}{\lambda_\varepsilon}
\nwc{\lam}{\lambda}
\nwc{\del}{\delta}
\nwc{\tht}{\theta}
\nwc{\om}{\omega}

\nwc{\btht}{\blds{\theta}}
\nwc{\bxi}{\blds{\xi}}
\nwc{\bmal}{\blds{\al}}
\nwc{\bmbe}{\blds{\beta}}

\nwc{\la}{\langle}    
\nwc{\ra}{\rangle}    
\nwc{\lp}{\left(}     
\nwc{\rp}{\right)}    

\nwc{\cc}{\overline}   

\nwc{\IA}{\mathbb{A}} 
\nwc{\IB}{\mathbb{B}} 
\nwc{\IC}{\mathbb{C}} 
\nwc{\ID}{\mathbb{D}} 
\nwc{\IE}{\mathbb{E}} 
\nwc{\IF}{\mathbb{F}} 
\nwc{\IG}{\mathbb{G}} 
\nwc{\IH}{\mathbb{H}} 
\nwc{\IN}{\mathbb{N}} 
\nwc{\IP}{\mathbb{P}} 
\nwc{\IQ}{\mathbb{Q}} 
\nwc{\IR}{\mathbb{R}} 
\nwc{\IS}{\mathbb{S}} 
\nwc{\IT}{\mathbb{T}} 
\nwc{\IZ}{\mathbb{Z}} 

\nwc{\ba}{\blds{a}}
\nwc{\bb}{\blds{b}}
\nwc{\bc}{\blds{c}}
\nwc{\bd}{\blds{d}}
\nwc{\be}{\blds{e}}
\nwc{\fb}{\blds{f}} 
\nwc{\bg}{\blds{g}}
\nwc{\bh}{\blds{h}}
\nwc{\bi}{\blds{i}}
\nwc{\bj}{\blds{j}}
\nwc{\bk}{\blds{k}}
\nwc{\bl}{\blds{l}}
\nwc{\bm}{\blds{m}}
\nwc{\bn}{\blds{n}}
\nwc{\bo}{\blds{o}}
\nwc{\bp}{\blds{p}}
\nwc{\bq}{\blds{q}}
\nwc{\br}{\blds{r}}
\nwc{\bs}{\blds{s}}
\nwc{\bt}{\blds{t}}
\nwc{\bu}{\blds{u}}
\nwc{\bv}{\blds{v}}
\nwc{\bw}{\blds{w}}
\nwc{\bx}{\blds{x}}
\nwc{\by}{\blds{y}}
\nwc{\bz}{\blds{z}}

\nwc{\bA}{\blds{A}}
\nwc{\bB}{\blds{B}}
\nwc{\bC}{\blds{C}}
\nwc{\bD}{\blds{D}}
\nwc{\bE}{\blds{E}}
\nwc{\bF}{\blds{F}}
\nwc{\bG}{\blds{G}}
\nwc{\bH}{\blds{H}}
\nwc{\bI}{\blds{I}}
\nwc{\bJ}{\blds{J}}
\nwc{\bK}{\blds{K}}
\nwc{\bL}{\blds{L}}
\nwc{\bM}{\blds{M}}
\nwc{\bN}{\blds{N}}
\nwc{\bO}{\blds{O}}
\nwc{\bP}{\blds{P}}
\nwc{\bQ}{\blds{Q}}
\nwc{\bR}{\blds{R}}
\nwc{\bS}{\blds{S}}
\nwc{\bT}{\blds{T}}
\nwc{\bU}{\blds{U}}
\nwc{\bV}{\blds{V}}
\nwc{\bW}{\blds{W}}
\nwc{\bX}{\blds{X}}
\nwc{\bY}{\blds{Y}}
\nwc{\bZ}{\blds{Z}}

\nwc{\va}{{\bf a}}
\nwc{\vb}{{\bf b}}
\nwc{\vc}{{\bf c}}
\nwc{\vd}{{\bf d}}
\nwc{\ve}{{\bf e}}
\nwc{\vf}{{\bf f}}
\nwc{\vg}{{\bf g}}
\nwc{\vh}{{\bf h}}
\nwc{\vi}{{\bf i}}
\nwc{\vj}{{\bf j}}
\nwc{\vk}{{\bf k}}
\nwc{\vl}{{\bf l}}
\nwc{\vm}{{\bf m}}
\nwc{\vn}{{\bf n}}
\nwc{\vo}{{\it o}}
\nwc{\vp}{{\bf p}}
\nwc{\vq}{{\bf q}}
\nwc{\vr}{{\bf r}}
\nwc{\vs}{{\bf s}}
\nwc{\vt}{{\bf t}}
\nwc{\vu}{{\bf u}}
\nwc{\vv}{{\bf v}}
\nwc{\vw}{{\bf w}}
\nwc{\vx}{{\bf x}}
\nwc{\vy}{{\bf y}}
\nwc{\vz}{{\bf z}}

\nwc{\vA}{{\bf A}}
\nwc{\vB}{{\bf B}}
\nwc{\vC}{{\bf C}}
\nwc{\vD}{{\bf D}}
\nwc{\vE}{{\bf E}}
\nwc{\vF}{{\bf F}}
\nwc{\vG}{{\bf G}}
\nwc{\vH}{{\bf H}}
\nwc{\vI}{{\bf I}}
\nwc{\vJ}{{\bf J}}
\nwc{\vK}{{\bf K}}
\nwc{\vL}{{\bf L}}
\nwc{\vM}{{\bf M}}
\nwc{\vN}{{\bf N}}
\nwc{\vO}{{\it O}}
\nwc{\vP}{{\bf P}}
\nwc{\vQ}{{\bf Q}}
\nwc{\vR}{{\bf R}}
\nwc{\vS}{{\bf S}}
\nwc{\vT}{{\bf T}}
\nwc{\vU}{{\bf U}}
\nwc{\vV}{{\bf V}}
\nwc{\vW}{{\bf W}}
\nwc{\vX}{{\bf X}}
\nwc{\vY}{{\bf Y}}
\nwc{\vZ}{{\bf Z}}

\nwc{\cA}{\ml{A}}
\nwc{\cB}{\ml{B}}
\nwc{\cC}{\ml{C}}
\nwc{\cD}{\ml{D}}
\nwc{\cE}{\ml{E}}
\nwc{\cF}{\ml{F}}
\nwc{\cG}{\ml{G}}
\nwc{\cH}{\ml{H}}
\nwc{\cI}{\ml{I}}
\nwc{\cJ}{\ml{J}}
\nwc{\cK}{\ml{K}}
\nwc{\cL}{\ml{L}}
\nwc{\cM}{\ml{M}}
\nwc{\cN}{\ml{N}}
\nwc{\cO}{\ml{O}}
\nwc{\cP}{\ml{P}}
\nwc{\cQ}{\ml{Q}}
\nwc{\cR}{\ml{R}}
\nwc{\cS}{\ml{S}}
\nwc{\cT}{\ml{T}}
\nwc{\cU}{\ml{U}}
\nwc{\cV}{\ml{V}}
\nwc{\cW}{\ml{W}}
\nwc{\cX}{\ml{X}}
\nwc{\cY}{\ml{Y}}
\nwc{\cZ}{\ml{Z}}

\nwc{\pa}{\partial}
\nwc{\Tr}{\textrm{Tr}}
\renewcommand{\qed}{\hfill$\blacksquare$}

\begin{document}

\title{Relaxation time of quantized toral maps.}

\author{Albert Fannjiang\dag,\  
St\'{e}phane Nonnenmacher\ddag\ and\ Lech Wo{\l}owski\dag$\ast$}

\thanks{\dag\ Department of Mathematics, 
University of California at Davis,
Davis, CA 95616, USA
({\tt fannjian@math.ucdavis.edu}).
The research of AF is supported in part by the grant from U.S. National
Science Foundation, DMS-9971322 and UC Davis Chancellor's Fellowship}
\thanks{\ddag\ Service de Physique Th\'eorique,
CEA/DSM/PhT (Unit\'e de recherche associ\'ee au CNRS) CEA/Saclay 91191
Gif-sur-Yvette c\'edex, France ({\tt nonnen@spht.saclay.cea.fr})}
\thanks{$\ast$\ Present address: School of Mathematics
University of Bristol, Bristol BS8 1TW, U.K ({\tt L.Wolowski@bristol.ac.uk})}

\begin{abstract}
We introduce the notion of the \emph{relaxation time} for noisy 
quantum maps on the $2d$-dimensional torus - a generalization of
previously studied \emph{dissipation time}. 
We show that relaxation time is sensitive to the
chaotic behavior of the corresponding classical system if one
simultaneously considers the semiclassical limit ($\hbar\to 0$) together with
the limit of small noise strength ($\ep\to 0$).

Focusing on quantized smooth Anosov maps, we exhibit a 
semiclassical r\'egime $\hbar<\ep^{E}\ll 1$ (where $E>1$) in which classical and quantum
relaxation times share the same
asymptotics: in this r\'egime, a quantized Anosov map relaxes to equilibrium fast, as
the classical map does. As an intermediate
result, we obtain rigorous estimates of the quantum-classical correspondence for noisy maps
on the torus, up to times logarithmic in $\hbar^{-1}$. On the other hand, we show that 
in the  ``quantum r\'egime'' $\ep\ll\hbar\ll 1$, quantum and classical relaxation
times behave very differently.
In the special case of ergodic toral symplectomorphisms 
(generalized ``Arnold's cat'' maps),
we obtain the exact asymptotics of the quantum relaxation time and precise 
the r\'egime of correspondence between quantum and classical relaxations.

\end{abstract}

\maketitle

\section{Introduction}

The notion of the dissipation time for classical systems has been introduced in various 
contexts in \cite{F2,F,FNW,FW} to study the
speed at which a conservative dynamical system converges to some equilibrium, 
when subjected to noise (e.g. due to interactions with 
the `environment').

In those references, the state of the system was represented by a probability density function, 
and the distance of the system from equilibrium was measured by the
mean-square fluctuations of the density w.r.to the equilibrium density. The term 
dissipation referred in those works to the process of the decay of 
density fluctuations during the noisy evolution.

In the present work we generalize our results to quantum-mechanical setting and introduce
the notion of the {\sl relaxation time}, which in the context of the above mentioned 
papers coincides exactly with the notion of the {\em dissipation time} and generalizes
it to the setting where relaxation of the system towards its equilibrium need not 
involve the presence of physical dissipation.
To uniformize the terminology, only the term {\em relaxation time} will
be used in the sequel. 
 
The relaxation time $\tau_c$ will now refer both to the time
scale after which the density fluctuations are reduced by
a fixed factor, and in general to the time scale on which the system finds itself 
in an intermediate state, roughly speaking, 'half-way' between the initial state and
the final equilibrium. 
 
The results obtained in \cite{FNW,FW} yielded the information about the asymptotic behavior of the 
relaxation time (in the limit when the noise strength $\eps$ tends to zero)
for a particular type of dynamics, namely volume-preserving maps
on a $d$-dimensional torus
phase space, for which the ``natural'' equilibrium density is the
constant function. 
Such torus maps constitute simple examples of dynamical systems with
proven chaotic behavior. 
Our main conclusion was that the asymptotic behavior of $\tau_c(\eps)$
strongly depends on the ergodic properties of the underlying
noiseless map. 
We found that the relaxation toward the equilibrium occurs much faster in the case 
of a chaotic dynamics, than for a ``regular'' one. 
More precisely, the relaxation
time displays two main behaviors in the small-$\ep$ limit: 

{\em Logarithmic-law} $\tau_c\sim \ln(\ep^{-1})$. In this case one speaks of 
{\sl fast relaxation} (short relaxation time). This behavior is characteristic of 
strongly chaotic systems,
e.g. maps with exponential mixing, including uniformly 
expanding or hyperbolic systems \cite{FNW}.
When the map is an (irreducible) linear
hyperbolic automorphism of the torus, 
the constant in front of the logarithm (the ``relaxation rate constant'')
can be computed explicitly,
and is related with the Kolmogorov-Sinai (KS) entropy of the map \cite{FW}.

{\em Power-law} $\tau_c\sim \eps^{-\beta}$. One then speaks of 
{\sl slow relaxation} (long relaxation time). This behavior virtually concerns all 
non-weakly-mixing systems (non-ergodic maps, Kronecker maps on the torus); it may 
also apply to systems with 
sufficiently slow (power-law) decay of correlations, like
intermittent maps \cite{B}. 

One can intuitively understand these opposite asymptotics through the way
the noiseless dynamics connects different spatial scales (or ``wavelengths'').
A chaotic map typically transforms modes of wavelength $\approx\ell$ into
modes of wavelength $\approx e^{\pm\lambda}\ell$, where $\lambda$ is the (largest) Lyapounov
exponent. By iteration, it will
transfer density fluctuations at scale $\ell$ into fluctuations at
scale $\ell'$ in a time $\sim |\log(\ell/\ell')|$. On the other hand, a noise of 
``strength'' $\ep$ strongly reduces fluctuations at wavelengths $\leq \ep$, acting 
effectively as a
ultraviolet cutoff. Therefore, $|\log\ep|$
is the minimal time needed for the system to bring fluctuations from all scales $1\geq\ell\geq \ep$
down to the scale $\ep$, where they get damped. 
On longer time scales the system can be thought of as in equilibrium.
On the opposite, a non-weakly-mixing system will mix different scales
at a much smaller speed, so fluctuations at wavelengths $\ell\gg\ep$ will take
a longer time to get damped. We believe that these various behaviors of the relaxation time
hold as well in the case of flows on compact phase spaces (the noise then
acts continuously in time, instead of ``stroboscopically'' for the case of maps \cite{Ki}).

In the present paper, we apply the notion of relaxation time to
{\em quantum} dynamical systems. To be able to use our ``classical'' results
of \cite{FNW}, we will focus on the quantum systems corresponding
to volume-preserving maps on the torus, namely quantized
maps on the torus. Besides being volume-preserving, the maps need
to be invertible and preserve the symplectic structure on the (necessarily even-dimensional) 
torus, that is, be {\em canonical}. Quantum maps have been
much studied in the last 25 years as convenient toy models of ``quantum chaos'' 
\cite{Haake,GraDE}.
According to the ``standard'' 
quantization schemes, compactness of the torus phase space 
leads to finite-dimensional quantum
Hilbert spaces, where the quantum maps takes the form of a unitary propagator.
Such finite-dimensional operators are obviously much easier to study numerically
than Schr\"odinger operators on $L^2(\IR^d)$. The semiclassical limit is recovered
when the dimension $N=(2\pi\hbar)^{-1}$ of the Hilbert space 
diverges. 

The influence of ``noise'' on an otherwise unitary 
quantum evolution has already attracted much attention, both in the mathematical 
\cite{lindblad}
and physics literature \cite{caldeira-leggett,gardiner,PZ}. Noise can be due to
interactions of the quantum system under study with uncontrolled 
degrees of freedom, like those of the
``environment'' of the system, or on the contrary internal degrees of freedom
not accounted for. The form of quantum noise we will consider is not the most general
one, it is obtained by quantizing the noise affecting the corresponding
classical system (section~\ref{s:qnoise}): the quantum equilibrium state is
then the fully mixed state with maximal Von Neumann entropy.
Several works have studied
the problem of relaxation in the framework of quantized maps, especially when the classical
dynamics is chaotic \cite{Br,fishman,manderfeld,BPS}. 
The effect of noise can be measured through various ways (growth of
the Von Neumann entropy, decay of purity, decay of ``fidelity'' etc.). One can
also observe how the {\sl spectrum} of the quantum noisy propagator
departs from unitarity \cite{Br,manderfeld,N,GS}; since the noisy propagator
is a non-normal operator, its spectral radius only influences
the {\it long-time} evolution of the system. On the opposite, the behavior for shorter times 
could possibly be analyzed through the {\it pseudospectrum} of the propagator \cite{davies}.
Our present study bypasses this spectral approach, by directly estimating 
the ``quantum relaxation time'' $\tau_q$: this quantity
indicates at which time the system has significantly relaxed to the equilibrium state, 
uniformly over all possible initial conditions. 

The problematic of quantum chaos (``where does a quantum system encode the
information that its classical limit is chaotic?'') yields another (more formal) reason to study 
the quantum relaxation time. Indeed, 
the above-described dichotomy between the two possible small-noise behaviors of
$\tau_c$ shows that the logarithmic-law is a decent indicator of chaotic
dynamics. Therefore, it seems reasonable to try using the small-$\ep$ behavior of
the quantum relaxation time
$\tau_q$ to characterize a quantum chaotic system. Yet, we are now
dealing with {\em two limits}: on the one hand, one expects the quantum system to mimic the
classical one only in the {\em semiclassical limit} $\hbar\to 0$; on the other hand, to
characterize the classical dynamics we
also want to consider the {\em small-noise limit} $\ep\to 0$. The major part of this article
will study the interplay between these two limits, which do not commute with each other.

In order to carry out
this program rigorously, we will focus our attention on a small subclass of the
maps studied in \cite{FNW}, namely the smooth Anosov
maps, which include the hyperbolic linear symplectomorphisms (or generalized
``Arnold's cat'' maps). As mentioned above, for such systems one can understand
the logarithmic behavior of the classical relaxation time through the ``mixing of scales''
performed by the dynamics. Quantum mechanics contains an intrinsic scale, namely
Planck's constant $\hbar$: it gives the size of the ``quantum mesh'' on the torus
which supports the Hilbert space (see section~\ref{s:setup}). This irreducible
scale allows one to estimate the {\em breaking time} for the quantum-classical 
correspondence, namely the time when the evolution (through the noiseless dynamics)
of quantum observable starts to strongly deviate from the evolution of the corresponding
classical observable (this time is often called {\em Ehrenfest time}, and we will
denote it by $\tau_E$) \cite{Zas,CC}. For a hyperbolic system, this time also
satisfies a logarithmic law $\tau_E\approx \frac{\ln(\hbar^{-1})}{\lambda}$,
which can be understood similarly as for $\tau_c(\ep)$:
$\tau_E$ is the shortest time needed for the system to transfer all scales 
$1\geq \ell\geq \hbar$
down to the ``quantum scale'' $\hbar$, where classical and quantum dynamics
depart from each other.

When switching on the noise, quantum and classical dynamics will also correspond to each other 
at least until the Ehrenfest time $\tau_E$, whatever the noise strength $\ep$. 
Therefore, if the classical system decays before the Ehrenfest time ($\tau_c<\tau_E$),
then the quantum system will decay around the same
time: $\tau_q\approx \tau_c$. This situation is described in Proposition~\ref{AQC} 
and Corollary~\ref{regime}.
This r\'egime was already studied in various semiclassical approaches to study convergence to
equilibrium in a quantum system subject to some type of noise
(see e.g. results regarding the spectrum of noisy quantum propagators \cite{Br,manderfeld,N,GS}, 
the rate of decoherence \cite{PZ,BPS,GSS} and its relation with quantum dynamical entropy
\cite{AF94,ALPZ,BCCFV}).

When one allows the noise strength to decrease together with Planck's constant,
the correspondence $\tau_q\approx\tau_c\sim \ln(\ep^{-1})$ remains valid as long as those times
are smaller than the breaking time $\tau_E$. Such a ``semiclassical r\'egime'' 
is partially analyzed in Section~\ref{s:uniform} for the case of smooth Anosov maps:
Theorem~\ref{t:anosov-uniform} identifies a condition of the form $\ep>\hbar^{1/E}$, 
which ensures that $\tau_q\approx\tau_c$
(the exponent $1/E<1$ depends on the expanding rates of the classical map). 
More precise estimates are obtained in Section~\ref{s:linear}
for the case of Anosov linear automorphisms of the torus.
Theorem~\ref{QDT} and Corollary~\ref{QC} state that 
the correspondence $\tau_q\approx\tau_c$ holds
under the milder condition $\ep\geq C \hbar$. 
One can check in this linear case that this condition ensures $\tau_{c}\leq\tau_{E}$,
which justifies the correspondence.
The correspondence between quantum and classical relaxation
times includes the prefactor in front of $\log(\ep^{-1})$. As mentioned above, 
this constant is related to the KS entropy of
the classical map, which also coincides with various types of 
\emph{quantum} dynamical entropies introduced in the algebraic quantization
schemes \cite{ALPZ,BCCFV}. 

In Section~\ref{s:quantum} we investigate 
the opposite situation (dubbed as the ``quantum limit'')
where the classical relaxation time is
longer than the Ehrenfest time. Beyond that time the quantum system will approach equilibrium
much slower than its classical counterpart, and rather independently of 
the noiseless dynamics. Precisely, we show in Proposition~\ref{Nfixed} that 
under the condition $\ep/\hbar\ll 1$ (meaning that the noise scale is smaller than
the quantum scale), the quantum relaxation time is bounded from below as  
$\tau_{q}\geq f(\hbar/\ep)$, where the function $f$ grows 
at a rate only depending on the ``shape'' of the noise.
In Remark~\ref{r:quantum},
we notice that a slightly stronger condition on the decay of $\ep/\hbar$ 
ensures that 
$\tau_q\gg\tau_{c}$ independently of unitary quantum dynamics. 
In such a r\'egime, the noise scale
is much smaller than the quantum mesh size, so the quantum evolution is insensitive
to the noise, and propagates almost unitarily. It is indeed irrelevant
to cutoff fluctuations at a scale $\ep$ when the smallest possible scale of
the system is $\hbar\gg\ep$.

As in the classical case, we believe that our results should extend to quantized
Anosov flows (for which exponential decay of correlations
has been recently proven in \cite{Liv}), like for instance the Laplace operator on 
a compact manifold of negative curvature.

To finish this section, we will compare our results on the
relaxation time with the related decay of \emph{fidelity}, which has recently received much
attention in the physics literature. Fidelity measures the discrepancy between, on the one hand,
the ``unperturbed'' evolution of an initial state $|\psi_o\ra$ under some quantum dynamics 
(say, a quantum map $U_N$, see section~\ref{q-maps}), on the other hand,
the evolution of the same initial state, but under a ``perturbed dynamics'' (say, the map
$U_N\,e^{-i2\pi N\eps Op_{N}(H)}$). The perturbing Hamiltonian $H$
is chosen randomly, but is \emph{independent of time}: this 
constitutes the major difference from our ``noise'', which is equivalent with
a random perturbation changing at each time step. The fidelity is then defined as 
$$
F(n)=\big|\la\psi_o|(U_N\,e^{-i2\pi N\eps Op_{N}(H)})^{-n}\;U_N^n|\psi_o\ra\big|^2\,.
$$
This quantity was first introduced in \cite{peres}, and several regimes of its decay have
been identified \cite{jalabert,prosen,beenakker,cerruti}, 
depending of the type of classical dynamics (chaotic vs. regular), and of the relative 
values of the
perturbation strength $\ep$ and Planck's constant $\hbar=(2\pi N)^{-1}$. 
In general, the fidelity starts to
decay around a certain ``fidelity time'' $n\approx \tau_F$, down to a saturation
where it oscillates
around values $\cO(\hbar)$. We will recall below
how $\tau_F$ depends on $\ep$ and $\hbar$ (when both are small),
in the case where the
classical dynamics is an Anosov map on the 2-dimensional torus, and
the initial state $|\psi_o\ra$
is a Gaussian wavepacket (coherent state) 
of width $\sqrt{\hbar}$.
We were able to identify at least four r\'egimes from the physics literature:
\begin{itemize}
\item for large enough perturbations, namely $\ep\gg \sqrt{\hbar}$, the fidelity decays
instantaneously, $\tau_F=1$.
\item in the range $\hbar\ll\ep\ll\sqrt{\hbar}$, the fidelity starts to decay at the time
$\tau_F\approx \frac{2\log(\ep^{-1})-(\log \hbar^{-1})}{2\lambda}$, which is comparable
with our ``log-time decay''.
\item for $\hbar^{3/2}\ll\ep\ll\hbar$, we are in the ``Fermi golden rule r\'egime'', and
$\tau_F\sim \big(\frac{\hbar}{\ep}\big)^2$.
\item $\ep\ll\hbar^{3/2}$ corresponds to the ``perturbative r\'egime'', where
$\tau_F\sim \frac{\sqrt{\hbar}}{\ep}$.
\end{itemize}
Subsequent r\'egimes are connected through crossovers, some of which
have been analyzed \cite{cerruti}.
The two last r\'egimes of weak perturbations are analog with our ``quantum limit'' for
the relaxation time. In these r\'egimes, the fidelity time is much longer than the 
Ehrenfest time $\tau_E$. Around $\tau_F$, the initial wavepacket is then
spread across the full torus, looking like a ``random state''; the same decay occurs
if we take for $|\psi_o\ra$ an arbitrary state.

In the first two r\'egimes of strong perturbation , the fidelity time satisfies 
$\tau_F\lesssim\frac{\tau_E}{2}$; therefore,
an evolved  coherent state is still localized in phase space around $\tau_F$. This 
shows that in these r\'egimes, the decay of fidelity crucially depends on the
choice for $|\psi_o\ra$ of an $\sqrt{\hbar}$-localized wavepacket. 
The inequality $\tau_F\leq\frac{\tau_E}{2}$ implies that
the quantum-classical correspondence still holds at the time $\tau_F$: this time
is thus asymptotically equal to the ``classical fidelity time'', 
which is the time when an initial classical density of width $\sqrt{\hbar}$,
evolved by the perturbed classical dynamics, departs from the same density
evolved by the unperturbed dynamics. Because the classical fidelity instantaneously decays 
for strong perturbations (as opposed to the logarithmic law for the classical relaxation time),
the quantum fidelity time $\tau_F$ does so too, thus behaving differently from the 
quantum relaxation time $\tau_q$.


\section{Setup and notation\label{s:setup}}

In all that follows, we use the following conventions to compare asymptotic behaviors of 
two quantities, for instance $a(\ep)$ and $b(\ep)$ in the limit $\ep\to 0$:
\begin{itemize}
\item $a(\ep)\ll b(\ep)$ iff $\frac{a(\ep)}{b(\ep)}\to 0$.
\item $a(\ep)\lesssim b(\ep)$ iff there is a constant $C>0$ such that $\frac{a(\ep)}{b(\ep)}\leq C$.
\item $a(\ep)\sim b(\ep)$ iff there are constants $C\geq c>0$ such that 
$c\leq \frac{a(\ep)}{b(\ep)}\leq C$.
\item $a(\ep)\approx b(\ep)$ iff $\frac{a(\ep)}{b(\ep)}\to 1$.
\end{itemize}


\subsection{Quantization on the Torus} $ $

The quantization on $\IT^{2d}$ presented below strictly follows that considered
in \cite{HB} and \cite{DE} in the $d=1$ case. The generalization
to arbitrary $d$ is in most aspects straightforward, and has been 
presented, in a slightly different notational setting, in \cite{Z,RSO,BonDB2}.  

\subsubsection{State Space and Observables} $ $
\label{q-algebra}
  
Let $T_{\bv}=e^{\frac{i}{\hbar} \bv \wedge \bZ}$ denote the standard Weyl
translation operators on $L^{2}(\IR^d)$, with $\bv=(\bq,\bp)\in \IR^{2d}$,
$\bZ=(\bQ,\bP)$ and $\bv\wedge\bZ= \bp\cdot \bQ - \bq\cdot \bP$.
Here $\bQ=(Q_{1},...,Q_{d})$ and $\bP=(P_{1},...,P_{d})$ 
denote the quantum position and momentum operators,
i.e. $Q_{j}\psi(\bx)=x_{j} \psi(\bx)$, 
$P_{j}\psi(\bx)=-i\hbar \pa_{x_{j}}\psi(\bx)$.

To quantize the torus, one extends the domain of $T_{\bv}$ to the space of
tempered distributions $\cS'(\IR^d)$, and considers its action on the
$\btht$-quasiperiodic elements (wavefunctions) of $\cS'(\IR^d)$, that is
distributions $\psi(\bq)$ satisfying:
\bea\label{quasiperiod}
\psi(\bq+\bm_{1})=e^{2\pi i \btht_{p}\cdot \bm_{1}}\psi(\bq), \qquad
(\ml{F}_{h}\psi)(\bp+\bm_{2})=e^{-2\pi i \btht_{q}\cdot \bm_{2}}(\ml{F}_{h}\psi)(\bp).
\eea
Here, the ``Bloch angle'' $\btht=(\btht_{q},\btht_{p})\in\IT^{2d}$ is fixed, while
$\bm=(\bm_{1},\bm_{2})$ takes any value in $\IZ^{2d}$. 
$\ml{F}_{h}$ denotes the usual quantum Fourier transform 
\bean
\label{CQFT}
(\ml{F}_{h}\psi)(\bp)=
\frac{1}{(2\pi\hbar)^{d/2}}\int_{\IR^{d}}\psi(\bq)e^{-i\frac{\bq \cdot \bp}{\hbar}}d\bq.
\eean
For any angle $\btht$, the space of such quasiperiodic distributions
is nontrivial iff $2\pi\hbar=h=1/N$ for a certain $N\in \IZ_{+}$.
From now on we only consider such values of Planck's constant. 
The corresponding space of wavefunctions
will be denoted by $\cH_{N}(\btht)$. It
forms a finite dimensional subspace of $\cS'(\IR^d)$ and can be 
identified with $\IC^{N^d}$.
The quasiperiodicity conditions \eqref{quasiperiod} can be restated in terms
of the action of translation operators:
\bea
\label{q-P}
\psi\in  \cH_{N}(\btht) \qquad \Leftrightarrow \quad \forall  \bm \in \IZ^{2d}, \qquad
T_{\bm}\psi=e^{2\pi i \lp \frac{N}{2} \bm_{1} \cdot \bm_{2} + \bm\wedge \btht \rp }\psi.
\eea 
That is, $\cH_{N}(\btht)$ consists of simultaneous eigenstates of all  
translations on the $\IZ^{2d}$ lattice.

A translation $T_{\bv}$ acts inside $\cH_{N}(\btht)$ iff $\bv\in N^{-1}\IZ^{2d}$, and
a natural Hermitian structure can be set on $\cH_{N}(\btht)$ such that all these
operators act unitarily.
This observation motivates the introduction of microscopic quantum
translations on $\cH_{N}(\btht)$:
$$
W_{\bk}=W_{\bk}(N,\btht):=T_{\bk/N |\cH_{N}(\btht)}=\big(e^{2\pi i \bk\wedge\bZ}\big)_{|\cH_{N}(\btht)}.
$$
The operators $W_{\bk}$ are indexed by points $\bk$ on the ``Fourier'' or ``reciprocal'' lattice $\IZ^{2d}$. 
Since they quantize the classical Fourier modes
$w_{\bk}(\bx)=e^{2\pi i \bk\wedge \bx}$,
they 
can be thought of as {\em Quantum Fourier Modes}.
The canonical commutation relations (CCR) take the form
\bea
\label{MCCR}
W_{\bk}W_{\bm}= e^{\frac{\pi i}{N}  \bk\wedge \bm}W_{\bk+\bm}, \qquad
W_{\bk}W_{\bm}= e^{\frac{2\pi i}{N} \bk\wedge \bm}W_{\bm}W_{\bk}.
\eea
Furthermore, the quasiperiodicity of the elements of
$\cH_{N}(\btht)$ induces a quasiperiodicity of the Quantum Fourier Modes
acting on that space.
Namely, for any $\bm\in\IZ^{2d}$ we have
\bea
\label{Wq-P}
W_{\bk +N\bm}(N,\btht)=e^{2\pi i \al(\bk,\bm,\btht)}\,W_{\bk}(N,\btht),
\eea
with the phase
\bean
\al(\bk,\bm,\btht)=\frac{1}{2}\bk\wedge\bm + \frac{N}{2} \bm_{1}\cdot\bm_{2} + \bm\wedge\btht.
\eean
The algebra of observables on the quantum space $\cH_{N}(\btht)$ is generated 
by the set of operators $\{W_{\bk}(N,\btht)\}_{\bk\in\IZ^{2d}}$ and will be denoted by $\cA_{N}(\btht)$. 
Due to quasiperiodicity, $\cA_{N}(\btht)$
is finite dimensional and can be identified (as a linear space) with the set of matrices
$\cL(\cH_{N}(\btht)) \cong \cM_{N^{d}\times N^{d}} \cong \IC^{N^{2d}}$. 

We select a fundamental domain $\IZ_{N}^{2d}$ of the quantum 
Fourier lattice. The choice centered around the origin 
seems to be the most natural one for our purposes (cf.~\cite{N}). Namely, we take for
fundamental domain the set of lattice points $\bk=(k_{1},...,k_{2d})\in \IZ^{2d}$ such that
\bean
\forall j\in \{1,...,2d\},\qquad k_{j}\in
\begin{cases}
\{-N/2+1,...,N/2\}, & \qquad \text{for $N$ even} \\ 
\{-(N-1)/2+1,...,(N-1)/2\}, & \qquad \text{for $N$ odd}.
\end{cases}
\eean
The set $\{W_{\bk}(N,\btht),\ \bk\in \IZ_{N}^{2d}\}$ forms
a basis for $\cA_{N}(\btht)$. Using the tracial state 
$\tau(A):=N^{-d} \Tr(A)$ on this algebra of matrices, we induce 
the Hilbert-Schmidt scalar product
\bean
\la A, B \ra = \tau(A^{*}B), \qquad A,B \in \cA_{N}(\btht).
\eean
The corresponding norm will be denoted by $\|\cdot\|_{HS}$. Equipped with this
norm, the above basis is orthonormal. 
One needs to keep in mind that $\|\cdot\|_{HS}$ does not coincide with
the standard operator norm, hence $\cA_{N}(\btht)$ is not considered here as a $C^{*}$-algebra.

\medskip

We can now easily quantize classical observables on $\IT^{2d}$. 
To any smooth observable $f\in C^{\infty}(\IT^{2d})$ with Fourier expansion
$f=\sum_{\bk\in \IZ^{2d}}\hat{f}(\bk)\,w_{\bk}$, corresponds
an element of $\cA_{N}(\btht)$, called its {\sl Weyl quantization}, denoted by $Op_{N,\btht}(f)$, and
defined as:
\bea\label{e:quantization}
Op_{N,\btht}(f)=\sum_{\bk \in \IZ^{2d}}\hat{f}(\bk)\;W_{\bk}(N,\btht)=\sum_{\bk \in \IZ_{N}^{2d}} 
\lp \sum_{\bm \in \IZ^{2d}}e^{2\pi i \al(\bk,\bm,\btht)}\hat{f}(\bk+N\bm) \rp W_{\bk}(N,\btht).
\eea
This quantization can be extended to observables 
$f\in L^{2}(\IT^{2d})$ satisfying $\sum_{\bk}|\hat{f}(\bk)|<\infty$.
 
The map $Op_{N,\btht}:C^\infty(\IT^{2d})\to \cA_{N}(\btht)$ is not injective. One can 
nevertheless define an {\sl isometric} embedding $W^{P}:\cA_{N}(\btht) \mapsto  L^{2}(\IT^{2d})$, which
associates with each quantum observable $A\in\cA_{N}(\btht)$ its polynomial Weyl symbol \cite{DEGI}
\bea\label{e:decompo-quant}
A=\sum_{\bk\in\IZ_{N}^{2d}} a_{\bk}\;W_{\bk}(N,\btht)\mapsto 
W^{P}(A)=\sum_{\bk \in \IZ_{N}^{2d}}a_{\bk}\;w_{\bk}.
\eea
The range of $W^{P}$ is the subspace $\cI_{N}=Span_{\IC}\{w_{\bk},\ \bk\in \IZ_{N}^{2d}\}$. The 
quantization map
$Op_{N,\btht}$ restricted to $\cI_N$ is the inverse of $W^P$.

The choice to work with the Hilbert structure on $\cA_{N}(\btht)$ corresponds to the
choice made in the classical setting to measure classical observables through their $L^2$ norm, 
rather than their $L^\infty$ norm.
With this choice, the notion of classical
relaxation (dissipation) time \cite{FNW, FW} can be straightforwardly extended to 
the quantum dynamics, and is suitable for semiclassical analysis.


\subsection{Quantization of toral maps\label{q-maps}} $ $

Let $\Phi$ denote a canonical map on $\IT^{2d}$, more precisely a $C^\infty$ diffeomorphism
preserving the symplectic form $\sum_j dp_j\wedge dq_j$. Any such map can be
decomposed into the product of three maps: 
$$
\Phi=F\circ t_{\bv} \circ \Phi_{1},
$$
where $F\in SL(2d,\IZ)$ is a linear automorphism of the torus, 
$t_{\bv}$ denotes the translation $t_{\bv}(\bx)= \bx+\bv$, and the function 
$\Phi_{1}(\bx)-\bx$ is periodic and has zero mean on the torus.

We will assume that the canonical map $\Phi_{1}$ is the
time-1 flow map associated with a Hamiltonian function on $\IT^{2d}$ (this Hamiltonian may
depend on time). In the case
$d=1$, this assumption is automatically satisfied \cite{CZ}.

To quantize $\Phi$, one first quantizes 
$F$, $t_{\bv}$ and $\Phi_{1}$ separately on $\cH_N(\btht)$. 
The quantization
of $\Phi$ is then defined as a composition of corresponding quantum maps
$U(\Phi)=U(F)\, U(t_{\bv})\, U(\Phi_1)$ \cite{KMR}. 
To each quantum map $U(\Phi)$ on $\cH_{N}(\btht)$ there
corresponds a quantum Koopman operator $\cU(\Phi)=\cU_{N,\btht}(\Phi)$ 
acting on $\cA_{N}(\btht)$ through the adjoint map 
$$
\cA_{N}(\btht)\ni A\mapsto \cU(\Phi)\, A=ad(U(\Phi))\,A=U(\Phi)^*A\, U(\Phi).
$$
In the next subsections we describe the quantizations of $F$, $t_{\bv}$ and $\Phi_{1}$
in some detail. 
The quantization procedure will ensure that the
correspondence principle holds. In our case this is expressed
by the Egorov property, which states that for  every 
$f\in C^{\infty}(\IT^{2d})$ there exists $C_f>0$ such that for any
angle $\btht$ and large enough $N$, 
\bea
\label{Egorov}
\|\cU_{N,\btht}(\Phi)\, Op_{N,\btht}(f) -Op_{N,\btht}(f\circ \Phi)\|_{HS}\leq \frac{C_f}{N}.
\eea
A more explicit estimate of the remainder is given in Proposition~\ref{p:Egorov}.

\subsubsection{Quantization of toral automorphisms} $ $
\label{catmap}

The symplectic map $F\in SL(2d,\IZ)$ acts on the algebra of
observables by means of its Koopman operator $K_{F}f=f\circ F$.
In the basis $\{w_{\bk}\}$ of classical Fourier modes, this operator acts 
as a permutation:
$K_{F}w_{\bk}=w_{F^{-1}\bk}$.
To define the quantum counterpart of this dynamics, we will bypass the
description of the quantum map $U(F)$ on $\cH_N(\btht)$, and directly
construct 
the quantum Koopman operator $\cU_{N,\btht}(F)$ acting on $\cA_{N}(\btht)$:
\bea
\label{QCatMap}
\cU_{N,\btht}(F)\,W_{\bk}=W_{F^{-1}\bk}.
\eea
For the dynamics to be well defined, $\cU_{N,\btht}(F)$ has to be a $^*$-automorphism of 
$\cA_{N}(\btht)$, i.e. its action 
must be consistent with the algebraic (CCR) and
quasiperiodic structures. The map $F$ is called quantizable, if for every 
$N$ there exist $\btht$ such that these consistency conditions are satisfied.
The appropriate condition can be formulated as follows (see \cite{HB,DE,RSO,BonDB2}):
\begin{prop}
\label{HBq}
A toral automorphism $F\in SL(2d,\IZ)$ is quantizable iff it is symplectic,
that is, $F\in Sp(2d,\IZ)$. 
For any given $N$, an angle $\btht$ is admissible iff it satisfies the following condition:  
\bea
\label{qbc}
\frac{N}{2}
\begin{pmatrix}
A \cdot B \\
C \cdot D
\end{pmatrix}
+F\btht=\btht \mod 1,
\eea
where $A,B,C,D$ denote block-matrix elements of F:
\bean
F=\begin{bmatrix}
A & B \\
C & D
\end{bmatrix}.
\eean
and $A\cdot B$ denotes the contraction of the
two matrices into a (column) vector:
\bean
(A\cdot B)_{i}=\sum_{j}A_{ij}B_{ij}.
\eean
\end{prop}
The existence of admissible angles is easy to establish.
If $N$ is even, one can simply choose $\btht=0$. This solution
can be chosen whenever all components of the vector 
$\lp
\begin{smallmatrix}
A \cdot B \\
C \cdot D
\end{smallmatrix}
\rp$
are even ('checkerboard' condition \cite{HB}).
Otherwise one considers two cases. 
If $F-I$ is invertible, then for any $\bk\in\IZ^{2d}$ the following angle is admissible:
\bean
\btht=(F-I)^{-1}\lp
\frac{N}{2}
\begin{pmatrix}
A \cdot B \\
C \cdot D
\end{pmatrix}
+\bk\rp.
\eean
This leads to $|\det(F-I)|$ distinct admissible angles.  
If $F-I$ is singular, one can construct an appropriate $\btht$ by applying the above considerations 
to the non-singular block.
We finally remark that in view of the defining condition (\ref{QCatMap}), the
Egorov property \eqref{Egorov} is
automatically satisfied (with no error term). 


\subsubsection{Quantization of a translation $t_{\bv}$\label{q-translations}} $ $

As explained in Section \ref{q-algebra}, a translation $t_{\bv}$ 
is quantized on $L^2(\IR^d)$ through a Weyl operator $T_{\bv}$. It was noticed
that such a quantum translation acts inside the algebra $\cA_{N}(\btht)$ only if
$\bv\in N^{-1}\IZ^{2d}$. In the opposite case, there are several possibilities to
quantize the translation \cite{BozDB}. We will choose the prescription given in \cite{MR}:
we take the vector $\bv^{(N)}\in N^{-1}\IZ^{2d}$ closest to $\bv$ (in Euclidean distance),
which can be obtained by taking, for each $j=1,\ldots,2d$,  the
component $v^{(N)}_j=\frac{[Nv_j]}{N}$,
where $[x]$ denotes the integer closest to $x$. 
One then quantizes $t_{\bv}$ on $\cH_{N}(\btht)$ through the restriction of $T_{\bv^{(N)}}$ on
that space (this is the same operator as $W_{[N\bv]}(N,\btht)$). 
The corresponding 
$^*$-automorphism on $\cA_{N}(\btht)$ is provided by $\cU_{N,\btht}(t_{\bv})=ad(T_{\bv^{(N)}})$.
The Egorov property \eqref{Egorov} holds for this quantization \cite{MR} 
(see also Appendix~\ref{a:Egorov}).


\subsubsection{Quantization of time-1 flow maps of periodic Hamiltonians} $ $

Let $\Phi_{1}$ denote the time-1 flow map associated with the periodic Hamiltonian $H(\bz,t)$,
meaning that $\Phi_t:\IT^{2d}\to\IT^{2d}$ satisfies the Hamilton equations:
\bean
\frac{\pa \Phi_{t}(\bz)}{\pa t}=\nabla^{\perp} H( \Phi_{t}(\bz),t),\qquad
\Phi_{0}=I.
\eean
To quantize $\Phi_{1}$, one applies the Weyl quantization to the Hamiltonian $H(t)$, 
obtaining a time-dependent Hermitian operator $Op_{N,\btht}(H(t))$. From there, one constructs
the time-1 quantum propagator on $\cH_{N}(\btht)$
associated with the Schr\"odinger
equation of Hamiltonian $Op_{N,\btht}(H(t))$: 
$$
U_{N,\btht}(\Phi_1):=\cT\, e^{-2\pi i N \int_{0}^{1} Op_{N,\btht}(H(t))\, dt}
$$
($\cT$ represents the time ordering).
As above, the corresponding $^*$-automorphism on  
$\cA_{N}(\btht)$ is defined as $\cU(\Phi_1)A=ad(U(\Phi_1))\,A$.
The Egorov property for such a propagator is proven in Appendix~\ref{a:Egorov}.


\subsection{Quantum Noise\label{s:qnoise}}

We briefly review the construction and properties of convolution-type 
noise operators in the classical setting. 
For more detailed description we refer to \cite{Ki,B,FNW}.
The construction starts with a continuous, even-parity probability density 
$g(\bx)\in L^1(\IR^{2d})$ representing the ``shape'' of the noise. This function
is sometimes assumed to be of higher regularity, and/or localized in a compact 
neighbourhood of the origin, and we will also require that $g(0)>0$. 
The noise strength (or magnitude) 
is then adjusted through a single parameter $\ep > 0$,
namely by defining the noise kernel using the rescaled density: 
\bean
g_{\ep}(\bx):=\frac{1}{\ep^{2d}}\;g\left(\frac{\bx}{\ep}\right)\ \mbox{on $\IR^{2d}$},\qquad
\tilde{g}_{\ep}(\bx):=\sum_{\bn\in \IZ^{2d}} g_{\ep}(\bx+\bn)\ \ \mbox{on $\IT^{2d}$}.
\eean
In the sequel we use the following notation for the Fourier transform on $\IR^{2d}$ and $\IT^{2d}$:
\begin{align}
\label{FT}
\forall \bxi\in\IR^{2d},\quad\hat{g}(\bxi)&:= \int_{\IR^{2d}}g(\bx)\,e^{-2\pi i \bxi\wedge \bx} d\bx\\
\forall\bk\in\IZ^{2d},\quad
\hat{\tilde{g}}(\bk)&:= \int_{\IT^{2d}}\tilde{g}(\bx)\,e^{-2\pi i \bk \wedge \bx} d\bx
=\la w_{\bk},{\tilde g}\ra.
\end{align}
One obviously has
$\hat{\tilde{g}}_{\ep}(\bk)=\hat{g}_{\ep}(\bk)=\hat{g}(\ep\bk)$. Therefore, the Fourier
expansion of ${\tilde g}$ reads
\begin{equation}
\tilde{g}_{\ep}(\bx)=\sum_{\bk\in \IZ^{2d}}\hat{g}(\ep\bk)\,w_{\bk}(\bx).
\end{equation}

The classical noise operator is defined on $L^2(\IR^{2d})\ni f$ as the convolution
$G_{\ep}f := \tilde{g}_{\ep}*f$. The Fourier modes $\{w_{\bk},\ \bk\in\IZ^{2d}\}$ 
form a basis of eigenvectors of $G_{\ep}$. The operator is compact,
self-adjoint and admits the following spectral decomposition
\bea
G_{\ep}f= \sum_{\bk\in \IZ^{2d}} \hat{g}(\ep\bk)\,\hat{f}(\bk)\; w_{\bk}.
\eea
For any noise strength $\ep>0$, the operator $G_{\ep}$ leaves invariant the constant density
(conservation of the total probability), but is strictly contracting 
on $L_{0}^{2}(\IT^{2d})$, the subspace of $L^2(\IT^{2d})$ orthogonal to the constant functions.
Using the parity of $g$, we notice that
$G_{\ep}$ can be represented as:
\bean
\label{nicerep}
G_{\ep}f 
=\int_{\IT^{2d}}\tilde{g}_{\ep}(\bv)\,K_{\bv}f\,d\bv,
\eean
where, $K_{\bv}$ is the Koopman operator associated with the translation $t_{\bv}$.

Using this formula, we can easily quantize the noise operator on $\cA_N(\btht)$ \cite{N}.
For this, we formally replace in the above integral the Koopman operator
$K_{\bv}$ by its quantization $\cU_{N,\btht}(t_{\bv})$ described in subsection \ref{q-translations}.
Since  $\cU_{N,\btht}(t_{\bv})$ is constant when $\bv$ varies on a ``cube'' of 
edges of length $\frac{1}{N}$, it is more convenient to adopt a 
different definition, and replace the above integral by a discrete sum, therefore defining the
quantum noise operator as:
\bean
\cG_{\ep,N,\btht}:=
\frac{1}{N^{2d}\,Z}\sum_{\bn\in \IZ_{N}^{2d}}\tilde{g}_{\ep} \lp\frac{\bn}{N}\rp 
\cU_{N,\btht}(t_{\bn/N})=
\frac{1}{N^{2d}\,Z}\sum_{\bn\in \IZ_{N}^{2d}}\tilde{g}_{\ep} \lp\frac{\bn}{N}\rp ad(W_{\bn}(N,\btht)).
\eean

We note that the assumption of continuity of $g$ is used in the above formula in an essential 
way. Indeed, the quantum noise operator depends only on a discrete set of values of $g$ (evaluated 
on the quantum lattice $\IZ^{2d}/N$) and cannot be unambiguously defined for a general $L^{1}$ density.

The role of the prefactor $\frac{1}{Z}$ is to ensure that $\cG_{\ep,N,\btht}$
preserves the trace (the quantum version of the classical conservation
of probability). One can easily check (see Appendix \ref{pqnoise}) that 
$Z=\tilde{g}_{\ep N}(0)$, which cannot vanish from our assumption $g(0)>0$. 
The spectrum of $\cG_{\ep,N,\btht}$ is similar to that
of its classical counterpart:
\begin{prop}
\label{qnoise}
$\cG_{\ep,N,\btht}$ admits as
eigenstates the Quantum Fourier modes $\{W_{\bk}(N,\btht),\ \bk\in\IZ_{N}^{2d}\}$, associated with
the eigenvalues 
\bea
\label{gamma}
\gamma_{\ep, N}(\bk) :=
\frac{\sum_{\bn\in\IZ^{2d}}g_{\ep N}(\bn)\,e^{-2\pi i \bk\wedge\bn/N}}
{\sum_{\bn\in\IZ^{2d}}g_{\ep N}(\bn)}.
\eea
\end{prop}

In the sequel we will often require higher regularity ($g\in C^{M}$ with $M\geq 1$) and fast decay 
properties of noise generating density $g$. In such cases we will often use the representation
of $\gamma_{\ep, N}$ obtained by applying the Poisson summation formula: 
\bea
\label{PSFgamma}
\gamma_{\ep, N}(\bk) = \frac{\sum_{\bm\in\IZ^{2d}}\hat{g}_{\ep N}\big(\frac{\bk}{N}+\bm\big)}
{\sum_{\bm\in\IZ^{2d}}\hat{g}_{\ep N}(\bm)}.
\eea

The conservation of
the trace is embodied in the fact that $\gamma_{\ep,N}({\bf 0})=1$.
Since the eigenvalues do not depend on the angle $\btht$, we will call the noise operator
$\cG_{\ep,N}$ from now on.
Let $\cA^{0}_{N}(\btht)$ be the space of observables of vanishing trace, that is
the quantum version of $L^2_0(\IT^{2d})$. We then 
introduce the following norm for operators acting on $\cA^{0}_{N}(\btht)$ (these are
sometimes called {\sl superoperators} in the physics literature):
\bea
\label{qnorm}
\|\cG_{\ep,N}\|:=\sup_{A\in \cA^{0}_{N}(\btht),\ \|A\|_{HS}=1} \|\cG_{\ep,N}A\|_{HS}.
\eea
Since $\cG_{\ep,N}$ is Hermitian, we get from its spectral decomposition
\bean
\|\cG_{\ep,N}\|=\max_{0\not=\bk\in \IZ_{N}^{2d}}\gamma_{\ep, N}(\bk).
\eean
The explicit formula for $\gamma_{\ep, N}(\bk)$, together with the fact that  $g(\bx)\geq 0$, 
show that the quantum noise operator acts as a strict contraction on 
$\cA^{0}_{N}(\btht)$ (if $g$ is compactly supported, strict contractivity
is guaranteed only for large enough  $\ep N$).


\subsection{Noisy quantum evolution operator and its relaxation time} $ $

For a given quantizable map $\Phi$ of the torus, 
we define the noisy quantum propagator 
by the composition \cite{BPS,GSS,N}
$$
\cT_{\ep,N}:=\cG_{\ep,N}\circ\cU_{N,\btht}(\Phi).
$$
This model assumes that noise is present at each step of the 
evolution, and acts as a memoryless Markov process.

We will also consider the  family of coarse-grained quantum propagators:
\bea
\label{coarse}
\tilde{\cT}_{\ep,N}^{(n)}&:=&\cG_{\ep,N}\circ\cU_{N,\btht}(\Phi)^{n}\circ\cG_{\ep,N}.
\eea
The latter type of dynamics assumes that some uncertainty is present 
at the initial and final steps (preparation and measurement of the system),
but not during the evolution. All these operators are trace-preserving, and
are strictly contracting on $\cA^{0}_{N}(\btht)$ (except for the case mentioned
at the end of section~\ref{s:qnoise}), but in general they are not normal
(their eigenstates are not orthogonal to each other).

We will study the action of these operators on the space $\cA^{0}_{N}(\btht)$, using
the norm \eqref{qnorm}. Mimicking the classical setting, 
we introduce the notion of quantum relaxation time associated with these two types 
of noisy dynamics:
\begin{equation}
\begin{split}
\label{ndiss}
\tau_{q}(\ep,N)&:=\min\{n \in \IZ_{+}:\|\cT_{\ep,N}^{n}\| < e^{-1}\},\\
\tilde{\tau}_{q}(\ep,N)&:=\min\{n \in \IZ_{+}:\|\tilde{\cT}_{\ep,N}^{(n)}\| < e^{-1}\}.
\end{split}
\end{equation}
As in the classical case, the relaxation time provides an intermediate
scale between the initial stage of the evolution (where the conservative dynamics
is little affected by
the noise) and the ``final'' stage when the noise has driven the system
to its equilibrium (an initial observable $A$ evolves towards
 $\tau(A)I$, which corresponds to a totally mixed state in the Schr\"odinger picture). 

In the remaining part of the paper we will analyze the behavior of the quantum relaxation 
time in various r\'egimes. To avoid any confusion we will reserve the
symbols $T_{\ep}$, $\tilde{T}_{\ep}^{(n)}$, $\tau_{c}(\ep)$, $\tilde{\tau}_{c}(\ep)$ 
for the corresponding propagators and times studied in \cite{FW,FNW}.


\section{Relaxation times in the ``quantum limit''}
\label{s:quantum}
The main goal of this section is the analysis of the relaxation time of 
noisy quantum maps on the torus, for fixed Planck's constant 
$h=N^{-1}$ and small noise strength $\ep$.
As we explained in 
Section \ref{q-maps}, the quantum Koopman operator $\cU_N(\Phi)$ on $\cA_N(\btht)$
associated with a canonical map $\Phi$
on the torus was constructed as the adjoint action of a unitary map
$U_N(\Phi)$ on $\cH_{N,\btht}$:
\bean
\cU_N(\Phi)A=ad\big(U_{N}(\Phi)\big)=U_{N}(\Phi)^{*}\,A\,U_{N}(\Phi), \qquad A\in \cA_{N}(\btht).
\eean
The unitary matrix $U_{N}(\Phi)$ admits an orthonormal basis of eigenfunctions 
$\psi_{k}^{(N)}\in \cH_{N}(\btht)$. Each projector  
$|\psi_{k}^{(N)}\ra\la\psi_{k}^{(N)}|$ is invariant through $\cU_N(\Psi)$. 
Therefore:
\begin{prop}
\label{qKoopman}
Any 
quantum Koopman operator $\cU_N$ on $\cA_{N}(\btht)$ admits unity in its spectrum,
with a degeneracy
at least $N^{d}$. As a consequence, for fixed $N$, 
the dynamics generated by $\cU_N$ on $\cA_{N}(\btht)$ is non-ergodic.
\end{prop}
In \cite[Corollary 3]{FNW}, we showed that the classical relaxation time 
behaves as a power-law in $\ep$ if the Koopman operator $K_{\Phi}$ has a 
nontrivial eigenfunction with a modicum of H\"older regularity. 
Although in the quantum setting the corresponding regularity assumption on
eigenstates of $\cU_{N}(\Phi)$ would be satisfied automatically 
(every observable is expressible as a finite combination of Fourier modes), 
one cannot apply this corollary directly here due to the different (discrete)
nature of the noise operator (cf. the remark ending this section). 
Nevertheless the main argument leading to the slow relaxation result is still valid.

\begin{prop}\label{Nfixed}
Assume that the noise generating density $g$ decays sufficiently fast at infinity: 
$\exists \gamma>2d$  s.t. $g(\bx)= \cO(|\bx|^{-\gamma})$ as $|\bx|\to\infty$ 
(resp. $g(\bx)=\pi^{-d}\exp(-\bx^2)$, resp. $g$ has compact support).

Then, for any angle $\btht$, and for any $\ep$, $N$,
the quantum noise operator on $\cA_N(\btht)$ satisfies
\bequ
\begin{split}
\label{estimate1}
\|1-\cG_{\ep,N}\|\leq C\,(\ep N)^\gamma,\qquad&\mbox{resp.}\qquad 
\|1-\cG_{\ep,N}\| \leq C\,e^{-\frac{1}{(\ep N)^2}},\\
\mbox{resp.}\qquad &\|1-\cG_{\ep,N}\|=0\qquad\mbox{if}\quad\ep N<1/C.
\end{split}
\eequ
All these bounds are meaningful in the limit $\ep N\ll 1$.
As a result, the quantum relaxation time 
associated with any quantized map $\cU_N(\Phi)$ is bounded as 
\bequ
\begin{split}
\tau_q(\ep,N)\geq C(\ep N)^{-\gamma},\qquad&\mbox{resp.}\qquad
C\,N^2\,e^{\frac{1}{(\ep N)^2}} \geq \tau_q(\ep,N)\geq c\;e^{\frac{1}{(\ep N)^2}},\\
\mbox{resp.}\qquad &\tau_q(\ep,N)=\infty\qquad\mbox{if}\quad\ep N<1/C.
\end{split}
\eequ 
The constants only depend on $g$, and are independent of the map $\Phi$.

Furthermore, for all these types of noise, there is a constant $\tilde c>0$ such that if
$\ep N<\tilde c$, the coarse-grained quantum dynamics does not undergo relaxation:
$\tilde{\tau}_{q}(\ep,N)= \infty$.
\end{prop}

\textbf{Proof.} We use the RHS of the explicit expression \eqref{gamma} for the eigenvalues 
$\gamma_{\ep, N}(\bk)$ of $\cG_{\ep,N}$. From the decay assumption on $g$, we see that
in the limit $\ep N\to 0$, 
\bea\label{e:sum_g}
\sum_{0\neq\bn\in\IZ^{2d}} g\bigg(\frac{\bn}{\ep N}\bigg)\leq 
C(\ep N)^\gamma\sum_{0\neq\bn\in\IZ^{2d}} \frac{1}{|\bn|^\gamma}.
\eea
The sum on the RHS converges because $\gamma>2d$. Therefore, we get
$0\leq 1-\gamma_{\ep, N}(\bk)\leq C(\ep N)^\gamma$ uniformly w.r.to $\bk\in \IZ_{N}^{2d}$.
Since $\cG_{\ep,N}$ is Hermitian, this yields the estimate \eqref{estimate1}. 

This implies that the noisy propagators contract very slowly, independently of the 
map $\Phi$:
\bean
\forall n\geq 0,\qquad\|\cT_{\ep,N}^{n}\| \geq \big(\min_{\bk\in\IZ^{2d}_{N}}  
\gamma_{\ep, N}(\bk)\big)^n
\geq \big(1-C(\ep N)^\gamma\big)^n,\qquad 
\|\tilde{\cT}_{\ep,N}^{(n)}\|\geq \big(1-C(\ep N)^\gamma\big)^2.
\eean
These inequalities prove the lower bound on $\tau_q$ in the case of
a power-law decay of $g$. If $g$ has compact support, the sum on the LHS of 
\eqref{e:sum_g} clearly vanishes if $\ep N$ is small enough, so that $\cG_{\ep,N}=1$ 
in this case.

The case of Gaussian noise is treated similarly, the LHS of Eq.~\ref{e:sum_g} being
clearly bounded above by $C\,e^{-1/(\ep N)^2}$. Besides, in that case
the largest $\gamma_{\ep, N}(\bk)$ 
(e.g. for $\bk=(1,0,\ldots,0)$) can be precisely estimated as $1-C\,N^{-2}\,e^{-1/(\ep N)^2}$, 
yielding the upper bound for $\tau_q(\ep,N)$.\qed

\begin{rem}\label{r:quantum}
In the case of Gaussian noise, we proved in \cite[Corollary 1]{FNW} that 
the classical relaxation time always satisfies the upper bound 
$\tau_{c}\lesssim \ep^{-2}$, independently of the map. 
Therefore, for this Gaussian noise, the bounds for $\tau_q$ obtained in the above Proposition show that
the quantum relaxation time is much larger than the classical one, regardless of the 
dynamics, as long as $\ep N \leq \frac{c}{\sqrt{\ln(\ep^{-1})}}$ for $c<1/\sqrt{2}$. 
In this r\'egime, the
noise width $\ep$ is smaller than the quantum mesh size $\sim\hbar$, therefore the
quantum dynamics does not feel the noise, and propagates (almost) unitarily. 
\end{rem} 


\section{Semiclassical analysis of the relaxation time}
\label{semi}

To extract information about the classical dynamics from the 
quantum relaxation time, one needs to consider a different r\'egime from the one
described in last section: what we need is a semiclassical r\'egime where Planck's constant
goes to zero together with the noise strength 
(cf. a similar discussion on the spectrum of $\cT_{\ep,N}$ in \cite[Section 5]{N}).

The semiclassical analysis relates the quantum and classical propagators to one another.
Following the notation introduced in Section~\ref{q-algebra}, for
any $N\in\IZ_+$  
we denote by $\Pi_{\cI^0_{N}}$ the orthogonal (Galerkin-type) projector of $L^2_{0}(\IT^{2d})$
onto its subspace $\cI^0_{N}=Span\{w_{\bk},\ \bk\in\IZ^{2d}_N- 0\}$. Using the fact that
$Op_N$ and its inverse $W^P$ realize isometric bijections between 
$\cI^0_N$ and $\cA^0_N$, to any operator $\cT_N\in\cB(\cA^0_{N}(\btht))$ we associate the operator 
\bean
\sigma_{N}(\cT_N):= W^{P}\cT_N Op_{N} \Pi_{\cI^0_{N}}
\eean
acting on $L^2_0(\IT^{2d})$. This operator is trivial on $(\cI^{0}_N)^{\perp}$, 
and its restriction on $\cI^0_N$ is isometric to $\cT_N$.
$\sigma_N$ therefore defines an isometric embedding of the finite dimensional algebra 
$\cB(\cA^{0}_{N}(\btht))$
into the infinite dimensional one $\cB(L^2_{0}(\IT^{2d}))$. 

It has been shown in \cite{N} (see Lemma~1 and its proof there)
that for any quantizable smooth map $\Phi$ and any fixed $\ep>0$, 
the operator $\sigma_{N}(\cT_{\ep,N})$ (isometric to $\cT_{\ep,N}=\cG_{\ep,N}\cU_N(\Phi)$)
converges in the limit $N\to\infty$ to the
classical noisy propagator $T_{\ep}=G_\ep\,K_\Phi$. This convergence
holds in the norm of bounded operators on $L^2_{0}(\IT^{2d})$.
 This implies in particular that for any fixed $\ep>0$ and $n\in \IN$
the sequence $\sigma_{N}(\cT^n_{\ep,N})$ converges to $T^{n}_{\ep}$ in the semiclassical limit.
The semiclassical 
convergence also holds for the coarse-grained propagators $\sigma_{N}(\tilde{\cT}^{(n)}_{\ep,N})$.
This convergence obviously implies the following behavior of the
quantum relaxation time:

\begin{prop}\label{AQC}
Let $\Phi$ be a smooth quantizable diffeomorphism on $\IT^{2d}$, and $g$ any noise generating density.
Then for any fixed noise strength $\ep>0$, 
the quantum relaxation time $\tau_q(\ep,N)$ (resp. $\tilde{\tau}_q(\ep,N)$) 
converges to the classical one $\tau_c(\ep)$ (resp. $\tilde{\tau}_c(\ep)$) in the
semiclassical limit.
\end{prop}
Using a standard diagonal argument, one obtains:

\begin{cor}\label{regime}
Under the conditions of the proposition, there exists a r\'egime $\ep\to 0$, $N(\ep)\to\infty$ 
such that $\tau_q(\ep,N(\eps))\approx \tau_c(\ep)$ 
(resp. $\tilde{\tau}_q(\ep,N(\eps))\approx \tilde{\tau}_c(\ep)$). 
Notice that these times necessarily diverge in this limit (cf.
Propositions 2 and 3 in \cite{FNW}).
\end{cor}

\textbf{Proof of the Proposition.}
We treat the case of the noisy relaxation times $\tau_c$ and $\tau_q$.
For given $\ep>0$, one has by definition
$\|T^{\tau_c}_{\ep}\|<e^{-1}$, $\|T^{\tau_c-2}_{\ep}\|> e^{-1}$ (the second inequality
is strict because $T_\ep$ is strictly contracting on $L^2_0(\IT^{2d})$). Therefore, the semiclassical
convergence of $\sigma_{N}(\cT^n_{\ep,N})$ towards $T_\ep$ implies the existence of an integer 
$N(\ep)$ such that for any $N\geq N(\ep)$, one
has simultaneously $\|\sigma_{N}(\cT^n_{\ep,N})^{\tau_c}\|<e^{-1}$ and 
$\|\sigma_{N}(\cT^n_{\ep,N})^{\tau_c-2}\|>e^{-1}$. This means that for $N\geq N(\ep)$, 
$\tau_q(\ep,N)=\tau_c(\ep)$ or $\tau_q(\ep,N)=\tau_c(\ep)-1$.

The proof concerning the coarse-graining relaxation time is identical. \qed

\medskip

Despite its generality, the above statement gives no information about the
behavior of the quantum relaxation time unless the behavior of the classical one is known. 
The latter has been investigated in \cite{FNW} for area-preserving maps on $\IT^{2d}$.
In particular, we have established logarithmic small-noise asymptotics 
$\tau_c(\eps)\sim \ln(\ep^{-1})$
(resp. $\tilde{\tau}_c(\eps)\sim \ln(\ep^{-1})$)
for a class of Anosov diffeomorphisms \cite[Theorem 4]{FNW}. 

Our aim in the next subsection is to apply these results and some of their refinements to obtain 
quantitative estimates on the semiclassical r\'egime for 
which quantum and classical relaxation times are of the same order.


\subsection{Uniform semiclassical r\'egimes\label{s:uniform}}$ $

In this section we derive an estimate on the growth of the function $N(\ep)$ for which the
classical-quantum correspondence of the relaxation times can be rigorously established.
To this end we derive and apply more precise Egorov estimates than the one expressed in Eq.~\eqref{Egorov}.
The main idea was already outlined in the Introduction:
for a generic map $\Phi$, the correspondence between classical and quantum
(noiseless) evolutions holds at least until the Ehrenfest time, the latter being of
order $|\log\hbar|$ if the map $\Phi$ is chaotic. Therefore, if the classical relaxation takes place
{\sl before} this Ehrenfest time, then the quantum relaxation should occur simultaneously
with the classical one. 

We will restrict ourselves to the case of Anosov maps on $\IT^{2d}$, which enjoy strong
mixing properties:

\begin{thm}\label{t:mixing}[Gou\"ezel-Liverani, \cite{GL}]
Let $\Phi$ be an Anosov $C^\infty$ diffeomorphism on $\IT^{2d}$, and let the noise
generating function $g$ be $C^\infty$ and compactly supported. Then, for any pair
of indices $s,s^*\in\IZ_+$ there exists $0<\sigma_{s,s^*}<1$ and $C>0$,
defining a function $\Gamma(n)=C\,\sigma_{s,s^*}^n$,
such that for small enough $\ep>0$, the correlations
between any pair of smooth observables $f$, $h$ with $\int f=0$ decay as follows:
\begin{equation}
\begin{split}
\forall n>0,\quad \Big|\int_{\IT^{2d}} f(\bx)\, h\circ \Phi^n(\bx)\,d\bx\Big|&\leq 
\Gamma(n)\,\|f\|_{C^{s_*}}\,\|h\|_{C^{s}},\\
\forall n>0,\quad
\Big|\int_{\IT^{2d}} f(\bx)\, T_\ep^n h(\bx)\,d\bx\Big|&\leq \Gamma(n)\,\|f\|_{C^{s_*}}\,\|h\|_{C^{s}}.
\end{split}
\label{e:clas-mixing}
\end{equation}
\end{thm}

This classical mixing allows us to slightly generalize our results of \cite{FNW}. In
particular, one does not need to assume any regularity condition on the invariant 
foliation of the map $\Phi$. The condition of compact 
support for the noise generating
kernel can probably be relaxed to functions $g$ in the Schwartz space $\cS(\IR^{2d})$ 
(C.~Liverani, private communication).

For such Anosov maps, we will exhibit a joint semiclassical r\'egime and small-noise
r\'egime, for which quantum and classical relaxation rates are similar.

\begin{thm}\label{t:anosov-uniform}
Let $\Phi$ be a quantizable Anosov $C^\infty$ diffeomorphism on $\IT^{2d}$, and
let the noise generating function $g$ be in the Schwartz space $\cS(\IR^{2d})$, so
that the classical correlations decay as in the previous theorem.

Then there exists an exponent $E=E(\Phi)$ such that in the r\'egime $\ep\to 0$,
$N=N(\ep) > \ep^{-E}$, the quantum relaxation times satisfy the same bounds
as their classical counterparts:

I) There exist $\tilde\Gamma>0$, $\tilde{C}>0$ 
such that the quantum coarse-grained relaxation time is bounded as:
\bean
  \frac{1}{\tilde\Gamma}\ln(\ep^{-1}) -\tilde{C} \leq
  \tilde{\tau}_q(\ep,N) \leq \frac{2d+s+s^*}{|\ln\sigma_{s,s^*}|} \ln(\ep^{-1}) + \tilde{C},
\eean
II) (Assume furthermore that the noise kernel $g$ is compactly supported.) 
There exists $\Gamma>0$, $C>0$ such that the quantum noisy
relaxation time satisfies:
\bean
\frac{1}{\Gamma}\ln(\ep^{-1}) - C
\leq  {\tau}_q(\ep,N) \leq \frac{2d+s+s^*}{|\ln\sigma_{s,s^*}|}\ln(\ep^{-1}) + C
\eean
\end{thm}
As mentioned above, the restriction to compactly-supported noise kernel in
statement (II) is probably unnecessary, so we put it into brackets.

The semiclassical r\'egime $N\ep^E>1$ of this Theorem
is quite distant from the ``quantum r\'egime'' ($N\ep\ll 1$) described in 
Proposition~\ref{Nfixed}. Inbetween we find a ``crossover range''
\bequ\label{e:cross}
\ep^{-1}\ll N \leq\ep^{-E}
\eequ
for which we do not control the quantum relaxation rates. However, at the level of characteristic
times, this range corresponds to differences between prefactors, as we summarize
in the following Corollary. There we define the ``Ehrenfest time''
precisely as $\tau_E=\frac{\ln(N)}{\Gamma}$, where $\Gamma$ is the largest expansion
rate of the Anosov map (see Lemma~\ref{l:chainrule}) instead of using the Lyapounov exponent 
$\lam$ (in general, $\lam$ and $\Gamma$ do not differ too much).

\begin{cor}\label{c:time-ranges} Assume the conditions of Theorem~\ref{t:anosov-uniform}.

i) In the semiclassical r\'egime $N \geq\ep^{-E}$, 
the Ehrenfest time is strictly larger than the classical and quantum relaxation times:
$$
\tau_E=\frac{\ln(N)}{\Gamma}\geq \frac{E\ln(\ep^{-1})}{\Gamma}\geq  \left\{
\begin{aligned}&K\,\tau_c(\ep)\\&K\,\tau_{q}(\ep,N)\end{aligned}\right.,
$$
with a constant $K>1$. 

ii) In the quantum r\'egime $N\ep\ll 1$, we have on the contrary
$$
\tau_E\leq \frac{\ln(\ep^{-1})}{\Gamma}\leq \tau_c(\ep).
$$

iii) For any $\gamma>0$, the noise kernel $g\in\cS(\IR^{2d})$ decays as $|\bx|^{-\gamma}$.
Then, in the ``deeply quantum'' r\'egime $N\ep\ll |\ln\ep|^{-1/\gamma}$
one has
$$
\tau_E\leq \frac{\ln(\ep^{-1})}{\Gamma}\leq \tau_c(\ep)\ll \tau_q(\ep,N).
$$
\end{cor}

This corollary is easily proven by using the bounds in the above theorem as well
as in its classical counterpart \cite[Th.~4 (II)]{FNW}, the explicit formulas
(\ref{e:E1},\ref{e:E3}) for the exponent $E$ and Proposition~\ref{Nfixed}.
It confirms the argument presented in the Introduction: the quantum
relaxation behaves like the classical one if both are shorter than the Ehrenfest time;
on the opposite, quantum relaxation becomes much slower than the classical one 
if the classical relaxation time is larger than $\tau_E$. 
Inbetween, the ``crossover range'' \eqref{e:cross} corresponds to a situation where
the classical relaxation time is of the same order as the Ehrenfest time, but where
we do not precisely control the quantum relaxation time.
 
\medskip

{\bf Remark:} 

The above theorem does only specify a r\'egime for which
the quantum and
classical relaxation times are {\sl of the same order}, 
$\tau_q(\ep,N(\ep))\sim \tau_c(\ep)\sim \ln(\ep^{-1})$. For a general
Anosov map $\Phi$ we are
unable to exhibit a r\'egime
for which $\tau_q(\ep,N(\ep))\approx \tau_c(\ep)$, that is for which the
relaxation times are {\sl asymptotic to each other} (cf. Corollary~\ref{regime}). 
The reason for this failure resides in
our insufficient knowledge of the 
observables which maximize the norms $\frac{\|T_\ep^n f\|}{\|f\|}$ 
(or $\frac{\|\tilde T_\ep^{(n)}f\|}{\|f\|}$).
These observables become quite singular
when $n$ becomes large, so we do not know whether
the quantum-classical correspondences stated in 
Propositions~\ref{p:Egorov}-\ref{p:Egorov-noisy} are helpful when
applied to these ``maximizing'' observables, if $n$ is close to the
classical relaxation time.

More precise estimates will be obtained in Section~\ref{s:linear} in the special case
of {\sl linear} Anosov diffeomorphisms of the torus.

\bigskip

{\bf Proof of Theorem \ref{t:anosov-uniform}:}

The proof will proceed in several steps. We start with 
refinements of the Egorov property \eqref{Egorov} for general maps $\Phi$. 
Then, we prove
lower bounds for the quantum relaxation times in the case of an expansive
map, and upper bounds if the map is mixing, so that both bounds can be
applied if $\Phi$ is Anosov.

\subsubsection{Egorov estimates}
The two following estimates (proven in Appendix~\ref{a:Egorov}) are 
obtained  
by adapting the methods of \cite{BouzRob} to quantum mechanics on $\IT^{2d}$.
To alleviate the notations we omit to indicate the dependence on the angle $\btht$.

\begin{prop}\label{p:Egorov}
Let $\Phi$ be a smooth quantizable map on $\IT^{2d}$, and $\cU_{N}(\Phi)$ its quantization
on $\cA_{N}$. Then there exists a constant $C>0$ such that
for any $N>0$, any classical observable $f\in C^\infty(\IT^{2d})$
and any $n\in\IN$, one has
\bea
\|\cU_{N}(\Phi)^n\, Op_{N}(f) -Op_{N}(f\circ \Phi^n)\|_{HS}\leq 
\frac{C}{N}\,\sum_{m=0}^{n-1}\|f\circ \Phi^{m}\|_{C^{2d+3}}.
\eea
\end{prop}

For a generic map $\Phi$, the  norm on the RHS will grow exponentially,
with a rate $e^{\Gamma n}$ where $\Gamma$ depends on the local hyperbolicity of
the map. For more ``regular'' maps,  the derivatives may
grow as a power law (cf. the discussion on the differential $D\Phi^n$ in
\cite[Section 4]{FNW}).

We will also need the following noisy version of the classical-quantum correspondence
(proven in Appendix \ref{a:Egorov-noisy}):

\begin{prop}\label{p:Egorov-noisy}
Assume that for some power $M\geq 2d+1$, the noise generating function $g\in C^M(\IR^{2d})$ and
all its derivatives up to order $M$ decay fast at infinity.

Let $\Phi$ be a quantizable map and $T_{\ep}$, $\cT_{\ep,N}$ the associated classical and
quantum noisy propagators. Then there exists $\tilde{C}>0$ such that,
for any $f\in C^\infty(\IT^{2d})$ and any $n\geq 0$,
\bea\label{e:Ego-noise}
\|\cT^n_{\ep,N}\, Op_{N}(f)-Op_{N}(T^n_\ep\,f)\|_{HS}\leq 
\tilde{C}\Big(\sum_{m=0}^{n-1}  \frac{\|T^m_\ep\,f\|_{C^{2d+3}}}{N}\Big)
+\tilde{C}\,\frac{\|T^n_\ep\,f\|_{C^M}}{(\ep N)^M},
\eea
where the implied constant depends only on $\Phi$ and $g$.
\end{prop}

Using these two propositions, we will now
to adapt the proofs given in \cite{FNW} for lower and upper bounds of 
the classical relaxation times, to the quantum framework. 


\subsubsection{Lower bounds for expansive maps}$ $

The lower bounds for the noisy relaxation time $\tau_c(\ep)$
rely on the following identity \cite[Section 4]{FNW}. 
Let $f$ be an arbitrary function in $C^1_0(\IT^{2d})$, e.g. the Fourier mode
$f=w_{\bk}$ for $\bk=(1,0,\ldots,0)$. For $g$ decaying fast at infinity,
we showed that for a certain $C>0$,
$$
\|T^n_\ep w_{\bk}\|_{L^2_0}\geq 1-C\ep \sum_{m=1}^n \|\nabla (T^m_\ep w_{\bk})\|_{L^2_0}
\geq 1-C\ep \|\nabla w_{\bk}\|_{C^0} \sum_{m=1}^n \|D\Phi\|^m_{C^0}.
$$
We will now use this formula to get a lower bound on the corresponding quantum quantity,
$\|\cT_{\ep,N}^n W_{\bk}\|_{HS}$. Indeed, from Eq.~\eqref{e:Ego-noise}, we have
for $M\geq 2d+3$:
\begin{equation}\label{e:lower-quant1}
\|\cT_{\ep,N}^n W_{\bk}\|_{HS}\geq 1-C\ep \|\nabla w_{\bk}\|_{C^0} \sum_{m=1}^n \|D\Phi\|^m_{C^0}
-\frac{C}{\min\big(N,(\ep N)^M\big)}\,\sum_{m=0}^{n}  \|T^n_\ep\,w_{\bk}\|_{C^M}.
\end{equation}
We need to control the
higher derivatives of $T^m_\ep w_{\bk}$. This can be done quite easily applying the 
chain rule (see \cite[Lemma 2.2]{BouzRob} and Appendix~\ref{a:chainrule}):
\begin{lem}\label{l:chainrule}
For any $C^\infty$ diffeomorphism $\Phi$, denote by
$\Gamma=\ln\big(\sup_x\|D\Phi_{|x}\|\big)$ the local expansion parameter of $\Phi$.
Then for any index $M\in \IN_0$, there exists a constant $C_M>0$ such that
$$
\forall f\in C^{\infty}(\IT^{2d}),\quad 
\forall n\geq 1,\qquad\|f\circ\Phi^n\|_{C^M}\leq C_M\,e^{nM\Gamma}\,\|f\|_{C^{M}}.
$$
Furthermore, for any $\ep>0$, the noisy evolution is also under control:
$$
\forall n\geq 1,\qquad\|T_\ep^n f\|_{C^M}\leq C_M\,e^{nM\Gamma}\,\|f\|_{C^{M}}.
$$
\end{lem}
We will only consider
the generic case of an expansive map, for which $\Gamma>0$. 
The inequality \eqref{e:lower-quant1} yields,
for $M\geq 2d+3$, the lower bound
\bea\label{low-noisy}
\|\cT_{\ep,N}^n\|\geq 1-C_M\big(\ep e^{n\Gamma} +
\big(N^{-1}+(\ep N)^{-M}\big)e^{nM\Gamma}\big).
\eea

The same lower bound can be obtained for the coarse-grained evolution. Indeed, 
$$
\tilde\cT_{\ep,N}^{(n)} W_{\bk}=\gamma_{\ep,N}(\bk)\,\tilde\cG_{\ep,N}\cU_{N}^n W_{\bk}.
$$
Using the Egorov estimate in Proposition~\ref{p:Egorov} and the bound \eqref{e:noise-corresp},
the norm of the RHS is bounded from below by
$$
|\gamma_{\ep,N}(\bk)|\,\Big(\|G_\ep w_{\bk}\circ\Phi^n\|_{L^2_0}
-C e^{nM\Gamma}(N^{-1}+(\ep N)^{-M})\Big).
$$
Since $g$ decays fast, the classical lower bound \cite[Eq.~(36)]{FNW} yields:
\begin{equation}\label{low-coarse}
\|\tilde\cT_{\ep,N}^{(n)}\|\geq 1-C\ep\|D\Phi^n\|_{C^0} -C e^{nM\Gamma}(N^{-1}+(\ep N)^{-M}),
\end{equation}
which is of the same type as the lower bound \eqref{low-noisy}. We assume that the derivative
of $\Phi^n$ grows with a rate $\tilde\Gamma>0$ (with $\tilde\Gamma\leq\Gamma$): there
is a constant $A>0$
such that for all $n>0$, $\|D\Phi^n\|_{C^0}\leq A\,e^{n\tilde\Gamma}$.

\begin{prop}\label{p:lower}
Assume that the noise generating function $g\in C^{M}$ with $M\geq 2d+3$, 
and all its derivatives decay fast at
infinity. For any smooth expansive diffeomorphism $\Phi$, we have in the joint limit 
$\ep\to 0$, $\ep N\to\infty$,
the following lower bounds for the quantum relaxation times:
\begin{align}
\tau_q(\ep,N)&\geq \min\Big(\frac{\ln (\eps^{-1})}{\Gamma},
\frac{\ln N}{M\Gamma},\frac{\ln(\ep N)}{\Gamma}\Big)
+C\\
\tilde\tau_q(\ep,N)&\geq \min\Big(\frac{\ln (\eps^{-1})}{\tilde\Gamma},
\frac{\ln N}{M\Gamma},\frac{\ln(\ep N)}{\Gamma}\Big)+C
\end{align}
\end{prop}
Since $M>2$, we conclude that in a r\'egime satisfying $N>\ep^{-M}$ (and respectively $N>\ep^{-\frac{\Gamma}{\tilde{\Gamma}}M}$)
, the above lower bounds for the quantum relaxation times are identical with the ones obtained for the
classical relaxation times.


\subsubsection{Upper bounds for mixing maps}$ $

In the classical framework \cite[Section 5]{FNW}, we used the Fourier decomposition
to get an upper bound on $\|T_\ep^n f\|$ for all possible $f\in L^2_0$, 
and then applied the classical mixing
(which holds for {\sl differentiable} observables) to the individual Fourier modes. 
Since our estimates of the quantum-classical correspondence 
(Props~\ref{p:Egorov},~\ref{p:Egorov-noisy})
apply to observables with some degree of differentiability,
this Fourier decomposition is well-adapted to the generalization
to the quantum framework.

Consider an arbitrary quantum observable $A\in\cA_N^0$, $\|A\|=1$ with Fourier coefficients
$\{a_{\bk}\}$. Using Fourier decomposition, we easily get for the coarse-grained evolution:
\begin{align}
\tilde\cT_{\ep,N}^{(n)} A&=\sum_{0\neq\bj\in\IZ_N^{2d}}\sum_{0\neq\bk\in\IZ_N^{2d}}
a_{\bk}\;\gamma_{\ep,N}(\bj)\gamma_{\ep,N}(\bk)
\langle W_{\bj},\cU^n_{N}(\Phi)W_{\bk}\rangle\,W_{\bj}\\
\Longrightarrow 
\|\tilde\cT_{\ep,N}^{(n)} A\|^2_{HS}&\leq 
\|A\|^2_{HS}\ \sum_{0\neq \bj,\bk\in\IZ_N^{2d}}|\gamma_{\ep,N}(\bj)\gamma_{\ep,N}(\bk)|^2\,
|\langle W_{\bj},\cU^n_{N}(\Phi)W_{\bk}\rangle|^2\label{e:upper1}
\end{align}
The overlaps $\la W_{\bj},\cU^n_{N}(\Phi)W_{\bk} \ra$ can be seen as quantum correlation functions.
From the Egorov estimate of Proposition~\ref{p:Egorov}, this correlation can be related to the
classical correlation function $\langle w_{\bj},w_{\bk}\circ\Phi^n\rangle$:
\begin{multline}
\langle W_{\bj},\cU^n_{N}(\Phi)W_{\bk}\rangle=\langle W_{\bj},Op_N(w_{\bk}\circ\Phi^n)\rangle +
\cO\Big(\frac{1}{N}\,\sum_{m=0}^{n-1}\|w_{\bk}\circ \Phi^{m}\|_{C^{2d+3}}\Big)\\
=\langle w_{\bj},w_{\bk}\circ\Phi^n\rangle+\sum_{0\neq \bm \in\IZ^{2d}}
(\pm)\langle w_{\bj+N\bm},w_{\bk}\circ\Phi^n\rangle 
+\cO\Big(\frac{1}{N}\,\sum_{m=0}^{n-1}\|w_{\bk}\circ \Phi^{m}\|_{C^{2d+3}}\Big).
\end{multline}
To write the second line, we used the explicit expression \eqref{e:quantization} for $Op_N(f)$. 
From the smoothness of $w_{\bk}\circ\Phi^n$, the 
sum over $\bm\neq 0$ on the RHS is an $\cO(N^{-M}\,\|w_{\bk}\circ \Phi^{n}\|_{C^{M}})$ 
for any $M>2d$. Therefore,
$$
\langle W_{\bj},\cU^n_{N}(\Phi)W_{\bk}\rangle=\langle w_{\bj},w_{\bk}\circ\Phi^n\rangle+
\cO\Big(\frac{1}{N}\,\sum_{m=0}^{n}\|w_{\bk}\circ \Phi^{m}\|_{C^{2d+3}}\Big).
$$
We can then use classical information on the derivatives of $w_{\bk}\circ \Phi^{m}$ and
the correlation functions $\langle w_{\bj},w_{\bk}\circ\Phi^n\rangle$. The former are estimated
in Lemma~\ref{l:chainrule}, while the latter depend on the dynamics generated by $\Phi$. 

\medskip

We now use the fact that the map $\Phi$ is mixing, both with and without noise,
in a way stated in Eqs.~\eqref{e:clas-mixing} (for a moment we do not need to
precise that $\Gamma(n)$ decays exponentially fast). 
Applied to the Fourier modes, Eqs.~(\ref{e:clas-mixing}) read (with $C$ depending only on
the indices $s,s^*$):
\begin{align}
\label{e:mixing}
\forall \bj,\bk\in\IZ^{2d}-0,\quad \forall n\in\IN,&\qquad
|\langle w_{\bj},w_{\bk}\circ\Phi^n\rangle|\leq C\,|\bj|^s\,|\bk|^{s_*}\Gamma(n),\\
\mbox{for any small enough $\ep>0$ and any $n\in\IN$,}&\qquad
|\langle w_{\bj},T_\ep^n w_{\bk}\rangle|\leq C\,|\bj|^s\,|\bk|^{s_*}\Gamma(n).
\label{e:mixing-noisy}
\end{align}

From this classical mixing, the quantum correlation functions are bounded from above as:
\bea\label{e:quant-corr}
|\langle W_{\bj},\cU^n_{N}(\Phi)W_{\bk}\rangle|\leq  C\,|\bj|^s\,|\bk|^{s_*}\Gamma(n) 
+ C\frac{\big(e^{n\Gamma}\,|\bk|\big)^{2d+3}}{N}.
\eea
We are now in a position to estimate the two sums in the RHS of Eq.~\eqref{e:upper1}. From
the estimate \eqref{e:est-gamma} and the fast decay at infinity of $g$, we can approximate
sums over the quantum noise eigenvalues by integrals \cite[Lemma 4]{FNW}:
\bea\label{e:sum}
\sum_{0\neq \bj\in\IZ_N^{2d}}|\gamma_{\ep,N}(\bj)|^2\,|\bj|^{2s}=
\frac{1}{\ep^{2s+2d}}\Bigg(\int |\hat g(\bxi)|^2\,|\bxi|^{2s}\,d\bxi +\cO(\ep)
+\cO\big((\ep N)^{2d+2s-2D}\big)\Bigg).
\eea
The exponent $D$ is related to the smoothness of $g$, and should
satisfy $D \geq 2d+1$. We will also assume that $D>d+s$ to make the last 
remainder small.
The estimate \eqref{e:sum} can be used to control the other terms appearing when combining
Eqs~\eqref{e:upper1} and \eqref{e:quant-corr}. The index $s$ will be replaced by $s_*$, 
$0$ and $2d+3$ respectively. In all cases, we will assume that $D>d+index$. The same
methods can be applied to estimate the norm of the noisy evolution $\cT_{\ep,N}$
(assuming a classical mixing of the type \eqref{e:mixing-noisy}).

\begin{prop}\label{p:upper-quant}
Assume that the noiseless and noisy dynamics generated by the map $\Phi$
are mixing, as in Eqs.~(\ref{e:clas-mixing}). Then the quantum
coarse-grained and noisy propagators satisfy the following bounds, in the
joint limits $\ep\to 0$, $\ep N\to\infty$:
\begin{align}
\label{e:upper-coarse}
\|\tilde\cT_{\ep,N}^{(n)} \|^2&\lesssim \frac{\Gamma(n)^2}{\ep^{2(2d+s+s_*)}}
+\frac{e^{2n(2d+3)\Gamma}}{N^2\,\ep^{8d+6}},\\
\|\cT_{\ep,N}^n \|^2&\lesssim \frac{\Gamma(n)^2}{\ep^{2(2d+s+s_*)}}
+\frac{e^{2n(2d+3)\Gamma}}{N^2\,\ep^{8d+6}}
+\frac{e^{2nM\Gamma}}{N^{2M}\,\ep^{4d+4M}}.
\label{e:upper-noisy}
\end{align}
In the second line, the upper bound holds for any exponent $M\geq 2d+1$.
\end{prop}
The first term in the RHS of those two equations
is of purely classical origin, it is identical to
the classical upper bounds \cite[Th. 3]{FNW} (remember that
the dimension of the phase space is now $2d$).
 This term decreases according
to the function $\Gamma(n)$, that is, according to the speed of mixing.
On the opposite, the remaining terms, due to quantum effects,
grow exponentially in time. 

\medskip

{\bf End of the proof of the Theorem}

We are now in position to combine our results for lower and upper bounds, in the case
of a smooth Anosov diffeomorphism. 
Such a diffeomorphism is expansive, therefore
it admits positive expansion parameters $\Gamma\geq\tilde{\Gamma}>0$, as defined
in Lemma~\ref{l:chainrule} and before Proposition~\ref{p:lower}. From that
Proposition, the constant 
\bequ\label{e:E1}
E_1:=(2d+3)\frac{\Gamma}{\tilde\Gamma}
\eequ
is such that in the r\'egime $N>\ep^{-E_1}$, the lower bounds for the
quantum and classical times are identical (in case of fully noisy 
dynamics it suffices to take $E_{1}:=2d+3$).

On the other hand, from Theorem~\ref{t:mixing} the classical mixing is
exponential, with a rate $\sigma_{s,s^*}<1$. As a result, this theorem and
the analysis of \cite{FNW} imply that the {\sl classical} relaxation
 times $\tilde\tau_c(\ep)$ and $\tau_c(\ep)$ 
are bounded from above by
\bea\label{e:clas-up}
\tau_c(\ep),\ \tilde\tau_c(\ep)\leq \frac{2d+s+s^*}
{|\ln\sigma_{s,s^*}|}\ln(\ep^{-1})+const.
\eea
We set $M=2d+3$ in Proposition~\ref{p:upper-quant}, and insert the upper bound 
\eqref{e:clas-up} in the second and third terms in
the RHS of Eqs.~(\ref{e:upper-coarse},\ref{e:upper-noisy}): these terms are
then of respective orders $\cO\big((N\,\ep^{E_2})^{-2}\big)$
and $\cO\big((N\,\ep^{E_3})^{-2(2d+3)}\big)$, where
\bequ\label{e:E3}
E_2=\Gamma\;\frac{(2d+3)(2d+s+s^*)}{|\ln\sigma_{s,s^*}|}+3+4d, \quad
E_3=\Gamma\;\frac{2d+s+s^*}{|\ln\sigma_{s,s^*}|}+2+\frac{2d}{2d+3}.
\eequ
The second exponent is clearly smaller than the first one. Therefore,
in the r\'egime $N\gg\ep^{-E_2}\gg\ep^{-E_3}$, these two terms are $\ll 1$
when $n$ is smaller than the classical relaxation times. Therefore in
this r\'egime the
{\sl quantum} relaxation times $\tilde\tau_q(\ep,N)$,
$\tau_q(\ep,N)$ are also bounded from above by
the RHS of Eq.~\eqref{e:clas-up}.

Finally, for any power $E>\max(E_1 ,E_3)$, the condition $N>\ep^{-E}$ provides
the ``semiclassical r\'egime''. Note that the exponent $E$ is defined from
purely classical quantities related to the map $\Phi$. \qed


\subsection{Relaxation time of quantum toral symplectomorphisms\label{s:linear}} $ $

In this section we analyze the quantum relaxation times when the map $\Phi$ is
a quantizable symplectomorphism $F$ of the torus $\IT^{2d}$ (see subsection~\ref{catmap}). 
We will only restrict ourselves to the
case where the matrix $F$ is {\sl ergodic} (none of its
eigenvalues is a root of unity), and diagonalizable. Let us remind
some notations we used in the classical setting \cite{FW}. Diagonalizability of $F$ implies
that there exists a rational basis of $\IR^{2d}$ where $F$ takes the form
$diag(A_1,\ldots,A_r)$, where each block $A_j$ is a $d_j\times d_j$ rational matrix, the
characteristic polynomial of which is irreducible over $\IQ$. The eigenvalues of $A_j$ are denoted
by $\{\lambda_{j,k},\ k=1,\ldots,d_j\}$. 
We call $h_j=\sum_{|\lambda_{j,k}|>1}\log|\lambda_{j,k}|$ the Kolmogorov-Sinai (K-S) entropy
of the block $A_j$, and $\hat h_j=\frac{h_j}{d_j}$ its ``dimensionally-averaged K-S entropy''.
Finally, we associate to the full matrix $F$ the ``minimal dimensionally-averaged
K-S entropy'' 
\bea
\label{min-entropy}
\hat h=\min_{j=1,\ldots,r}\hat h_j.
\eea
Due to the simple action of the map $\cU_N(F)$ on the quantum Fourier modes (Eq.~\eqref{QCatMap}),
many computations can be carried out explicitly, and yield precise asymptotics of
the quantum relaxation times.

To focus attention and avoid unnecessary notational and computational complications,
we restrict the considerations of this subsection to an isotropic Gaussian noise 
$\hat{g}(\bk)=e^{-|\bk|^{2}}$ (in \cite{FW} a
slightly more general noise was considered, given by $\alpha$-stable laws). 
 
From the exact Egorov property \eqref{QCatMap} and the fact that the quantum Fourier modes
$W_{\bk}(N,\btht)$ are eigenstates of the
quantum noise operator (cf. Proposition \ref{qnoise}), 
one easily proves that any $A\in \cA^{0}_{N}(\btht)$ with Fourier coefficients $\{a_{\bk}\}$
(cf. Eq.~\ref{e:decompo-quant}) evolves into
\bean
\cT_{\ep,N}^{n}A=\sum_{0\neq\bk\in \IZ^{2d}_{N}} a_{\bk}\,
\Big(\prod_{l=1}^{n}\gamma_{\ep N}(F^{-l}\bk)\Big)\, W_{F^{-n}\bk}.
\eean
Orthogonality of the $\{W_{\bk}\}$ then induces the exact expression:
\bea
\label{nd}
\|\cT_{\ep,N}^{n}\|=\max_{0\not=\bk\in \IZ^{2d}_{N}}
\bigg(\prod_{l=1}^{n}\gamma_{\ep N}(F^{-l}\bk)\bigg)=
\max_{0\not=\bk\in \IZ^{2d}_{N}}
\bigg(\prod_{l=1}^{n}\gamma_{\ep N}(F^{l}\bk)\bigg),
\eea
Similarly, in the coarse grained case we have
\bea
\label{cnd}
\|\tilde{\cT}_{\ep,N}^{(n)}\|=\max_{0\not=\bk\in \IZ^{2d}_{N}}
\big(\gamma_{\ep N}(\bk)\gamma_{\ep N}(F^{n}\bk)\big).
\eea
Using these exact formulas, we can precisely estimate the
quantum relaxation times.
\begin{thm}
\label{QDT}
Let $F\in Sp(2d,\IZ)$ be ergodic and diagonalizable, and for all $N\in\IN$ we
select an admissible angle $\btht$ for which $F$ may be quantized on $\cH_{N,\btht}$. 
The noise is assumed to be Gaussian. Then the quantum relaxation times associated with
the quantum dynamics satisfy the following estimates: 

I) For any $\ep>0$ and $N\in\IZ_+$, 
\bean
\tau_{q}(\ep,N) \geq \tau_{c}(\ep),\qquad
\tilde{\tau}_{q}(\ep,N) \geq \tilde{\tau}_{c}(\ep).
\eean

II) There exists $M>0$ (made explicit in Eq. \eqref{M}) such that in the joint limit 
$\ep\to 0$, $N>M\ep^{-1}$,
\bean
\tau_{q}(\ep,N) \approx \tau_{c}(\ep) \approx \frac{1}{\hat{h}(F)}\ln(\ep^{-1})
\eean

III) Let $\mu=\max(\|F\|,\|F^{-1}\|)$. For any coefficient $\beta>\frac{\ln\mu}{2\hat{h}(F)}+1$,
one has in the joint limit $\ep \to 0$, $N>\ep^{-\beta}$:
\bean
\tilde{\tau}_{q}(\ep,N) &\approx& \tilde{\tau}_{c}(\ep)\approx \frac{1}{\hat{h}(F)}\ln(\ep^{-1}).
\eean

Here $\hat{h}(F)$ is the minimal dimensionally averaged K-S entropy 
of $F$, Eq.~\eqref{min-entropy}.

\end{thm}

As a direct corollary of the above theorem (and using Proposition~\ref{Nfixed}), 
we obtain the following relations between, on one
side, the ``spatial'' scales (namely $\ep$ for the noise,  $\hbar$ for the scale of the
``quantum mesh''), and on the other side the ``time scales'' (namely the relaxation and Ehrenfest
times), for the case of linear ergodic (diagonalizable) symplectomorphisms. 
 As in Corollary~\ref{c:time-ranges}, we take for the Ehrenfest 
time $\tau_E=\frac{\ln N}{\Gamma}$, with now $\Gamma=\ln(\|F\|)$.

\begin{cor}\label{QC}$ $

Under the assumptions of Theorem \ref{QDT} the following relations hold between the
noisy quantum relaxation time and the Ehrenfest time $\tau_E$, in the joint limit
$\ep\to 0$, $N\to\infty$, depending on the behavior of the product $\ep N$:

i) If $N \gg \ep^{-1}$, then 
$$
\tau_E\gtrsim \tau_{c}(\ep)\approx\tau_{q}(\ep,N).
$$
The first $\gtrsim$ can be replaced by $\geq$ if $N\gg \ep^{-\Gamma/\hat h(F)}$.

ii) There exists $M>0$ (see \eqref{M}) such that, for any finite $M'>M$,
\bean   
\mbox{if}\ \ \ \ep\,N \to M' \ \ \ \mbox{then}\ \ \ \ \tau_{c}(\ep)\approx\tau_{q}(\ep,N) \sim \tau_{E}.
\eean 

iii) If $\ep\, N \leq \frac{1-\delta}{\sqrt{\ln\ln(\ep^{-1})}}$ for some $\delta>0$, then
\bean
 \tau_{E}\leq \tau_{c}(\ep) \ll \tau_{q}(\ep,N).
\eean
\end{cor}
The form of the ``deeply quantum r\'egime'' {\it iii)} is due to the Gaussian noise (compare with
Corollary~\ref{c:time-ranges}{\it iii)} for a more general noise).
For linear automorphisms, the ``crossover range'' 
is much thinner than for a nonlinear Anosov map (see Corollary~\ref{c:time-ranges}):
here this crossover takes place when Planck's constant $N$ crosses
a window $[\frac{\ep^{-1}}{\sqrt{\ln\ln(\ep^{-1})}}, M\ep^{-1}]$, to be compared with
a window $[\frac{\ep^{-1}}{\sqrt{\ln\ln(\ep^{-1})}}, \ep^{-E}]$ for 
a general Anosov map with Gaussian noise.

\bigskip

\textbf{Proof of Theorem \ref{QDT}.}

To prove the theorem we will need the following estimates (proven in Appendix \ref{est}),
which relate the eigenvalues of the classical and quantum noise operators. 
We remind that here and below, $\hat{g}_\sigma(\bxi)=e^{-|\sigma\bxi|^2}$.

\begin{lem} 
\label{GES}
For any $N\in\IN_0$ and $\bxi\in \IR^{2d}$, we denote by $\bxi^N$ the unique vector
in $\IR^{2d}$ s.t. all its components satisfy
$\bxi^N_j\equiv\bxi_j \bmod N$ and  $\bxi^N_j\in (-N/2,N/2]$. 

Then for any $\ep>0$, $N\in\IN_0$ and all $\bxi\in \IR^{2d}$, 
\bea
\label{pGauss}
\hat{g}_{\ep} (\bxi) \leq \hat{g}_{\ep} (\bxi^N) \leq 
\gamma_{\ep,N}(\bxi) \leq 
\frac{\hat{g}_{\ep}(\bxi^N)}{\tilde{g}_{\ep N}(0)}
+4d\,e^{-\frac{(\ep N)^2}{4}}
\leq
\hat{g}_{\ep}(\bxi^N)+4d\,e^{-\frac{(\ep N)^2}{4}}.
\eea
\end{lem}

Besides, we will need the following integer programming result \cite{FW}, which 
measures the ``minimal extension'' of an $F$-trajectory on the Fourier lattice:

\begin{prop}
\label{thmart}
Let $F\in SL(2d,\IZ)$ be ergodic and diagonalizable. 
For any (small) $\del>0$, there exists $n(\del)>0$ s.t. for any
$n\geq n(\del)$, we have: 
\bea
\label{ulestimateA}
e^{2(1-\del)\hat{h}(F)n} <
\min_{0 \not=\bk \in \IZ^{2d}} \big(|\bk|^2+|F^{n}\bk|^{2}\big) <
\min_{0 \not=\bk \in \IZ^{2d}} \sum_{l=0}^{n}|F^{l}\bk|^{2} <
e^{2(1+\del)\hat{h}(F)n}
\eea
As above, $\hat{h}(F)$ is the minimal dimensionally-averaged entropy \eqref{min-entropy}.
\end{prop}

We start to prove the statement {\it I)} of the Theorem.
According to the explicit equations (\ref{nd},\ref{cnd}) and their classical counterparts
\cite{FW}, the norms of the noisy and coarse-grained propagators
are given in terms of products of coefficients $\gamma_{\ep, N}(\bk)$ 
(resp. coefficients $\hat{g}_\ep(\bk)$ for the classical propagators).
Lemma \ref{GES} shows that for any $\bk\in\IZ^{2d}$, 
$\gamma_{\ep, N}(\bk)\geq \hat{g}_\ep(\bk)$. Applying this inequality factor by factor
in the explicit expressions for classical and quantum norms yields:
\bean
\forall n\geq 1,\quad \|\cT_{\ep,N}^{n}\|\geq 
\|T_{\ep}^{n}\|,\qquad
\|\tilde{\cT}_{\ep,N}^{(n)}\|\geq
\|\tilde{T}_{\ep}^{(n)}\|,
\eean
which yield the statement {\it I)}.

\medskip

The lower bounds of statements {\it II)} and {\it III)}
follow from the general inequalities established in point {\it I)}, together with
small-noise results obtained in the classical setting \cite{FW}. 

To prove the upper bound of statement {\it III)},
we bound from above the RHS of Eq.~\eqref{cnd}.
Given a coefficient $\beta$ as in the statement, we fix some (arbitrarily small)  
$\del>0$ satisfying
$\beta-1>\frac{\ln\mu}{2(1-\del)\hat{h}}$ (from here on, we abbreviate $\hat{h}(F)$ by $\hat{h}$).
In the r\'egime $\ep^{\beta} N >1$, for sufficiently small $\ep>0$ there
exist integers $n$ in the interval
\bea
\label{Et}
\frac{1}{(1-\del)\hat{h}}\ln(2\ep^{-1})< n<n+1
<\frac{1}{(1-\del)\hat{h}+\frac{1}{2}\ln\mu}\ln (N/2).
\eea
We take $\ep$ small enough such that the LHS of this equation is larger than
the threshold $n(\delta)$ defined in Proposition~\ref{thmart}.
We want to control the product
$\gamma_{\ep N}(\bk_0)\gamma_{\ep N}(F^{n}\bk_0)$ for integers $n$ in this interval,
uniformly for all
$0\neq \bk_{0}\in \IZ^{2d}_{N}$. We need to consider two cases.

\medskip
\begin{itemize}
\item If both $\bk_{0}$ and $F^{n}\bk_{0}$ 
belong to the ``fundamental cell'' $\IZ^{2d}_{N}$, 
then from Proposition \ref{thmart}, we have
\bea
\label{classmin}
 |\bk_{0}|^{2}+|F^{n}\bk_{0}|^{2} \geq 
 \min_{0 \not = \bk \in \IZ^{2d}}(|\bk|^{2}+|F^{n}\bk|^{2})>e^{2(1-\del)\hat{h}n}.
\eea
Thus for any such $\bk_{0}$, 
$\max\big(|\bk_0|,\ |F^{n}\bk_{0}|\big)> \frac{1}{\sqrt{2}}\,e^{(1-\del)\hat{h}n}$.
Using (\ref{pGauss}) and the fact that all $\gamma_{\ep,N}(\bk)<1$, we obtain:
\bean
\gamma_{\ep,N}(\bk_0)\gamma_{\ep,N}(F^{n}\bk_{0})<
\min\big(\gamma_{\ep,N}(\bk_0),\gamma_{\ep,N}(F^{n}\bk_{0})\big)
\leq \exp\big\{-\frac{\ep^2}{2}\, e^{2(1-\del)\hat{h}n}\big\} + C\,e^{-\frac{(\ep N)^2}{4}}.
\eean
From the left inequality in \eqref{Et}, the argument of the exponential in the above RHS is
smaller than $-2$. Since 
$\ep N>\ep^{1-\beta}\gg 1$, the product on the LHS is $<e^{-1}$.

\item assume the opposite situation: $\bk_0\in \IZ^{2d}_{N}$ but its image 
$F^{n}\bk_{0} \not \in \IZ^{2d}_{N}$. In that case, we may assume that the set
$S_0=\{\bk_0,F\bk_0,\ldots,F^{l_0-1}\bk_0\}\subset\IZ^{2d}_{N}$, while 
$F^{l_0}\bk_0\not\in\IZ^{2d}_{N}$. Consider also $\bk_n =(F^{n}\bk_{0})^N$ the
representative of $F^{n}\bk_{0}$ in the fundamental cell, and assume that
$S_n=\{\bk_n,F^{-1}\bk_n,\ldots,F^{-l_n+1}\bk_n\}\subset\IZ^{2d}_{N}$, while 
$F^{-l_n}\bk_n\not\in\IZ^{2d}_{N}$. Obviously, the sets $S_0$, $S_n$ have no
common point (this would let the full trajectory $\{F^j\bk_0\}_{j=0}^n$ be contained
in $\IZ^{2d}_{N}$), so that $l_0+l_n\leq n+1$. The vectors $k_0$, $k_n$ satisfy
the obvious inequalities
\bean
\frac{N}{2}\leq|F^{l_0}\bk_{0}|\leq \|F\|^{l_0}\,|\bk_0|\leq \mu^{l_0}\,|\bk_0|,\qquad 
\frac{N}{2}\leq|F^{-l_n}\bk_{n}|\leq \|F^{-1}\|^{l_n}\,|\bk_n|\leq \mu^{l_n}\,|\bk_n|.
\eean 
Since $\min(l_0,l_n)\leq \frac{n+1}{2}$, either $|k_0|$ or $|k_n|$ is bounded from below
by $\frac{N}{2}\mu^{-\frac{n+1}{2}}$, and, from 
the right inequality in \eqref{Et}, 
also by $e^{(1-\del)\hat{h}n}$.
We are back to the lower bound of the previous case,
leading to the same conclusion.
\end{itemize}

We have therefore proven that for sufficiently small
$\ep>0$ and $N>\ep^{-\beta}$, any integer $n$ in the (nonempty) interval \eqref{Et} satisfies
$\|\tilde{\cT}_{\ep,N}^{(n)}\|<e^{-1}$,
and is therefore $\geq\tilde{\tau}_q(\ep,N)$. As a result,
\bean
\frac{1}{(1-\del)\hat{h}}\ln(2\ep^{-1})+1\geq \tilde{\tau}_q(\ep,N).
\eean
Since $\del$ can be taken arbitrarily small, we obtain the statement {\it III)} of the Theorem.

\bigskip

The upper bounds of statement {\it II)} is proven with similar methods. 
We want to bound from above the product \eqref{nd}.
Let $C$ denote the constant of the RHS of \eqref{pGauss}, and take $M=M(F)$ 
a constant such that
both conditions below are satisfied:
\bea
\label{M}
Ce^{-\frac{M^2}{4}}<e^{-2},\qquad 
\frac{1}{\hat{h}} \ln\bigg(\frac{M}{4\,\|F\|}\bigg)>2.
\eea
Let us fix some $0<\del'<\del<1/2$. If $\ep N > M$, the second condition implies the existence of 
an integer $n$ such that
\bea
\label{Et1}
\frac{1}{(1-\del)\hat{h}}\ln(2\,\ep^{-1})< n-1 
<\frac{1}{(1-\del)\hat{h}}\ln \lp \frac{N}{2\|F\|}\rp.
\eea

We take $\ep$ small enough so that any $n$ in the above interval is larger than the threshold
$n(\del')$ of Proposition~\ref{thmart}, and also satisfies
$e^{2(\del-\del')\hat{h}n}>n$. For such an $n$,  we can then estimate the products 
$\prod_{l=0}^{n-1}\gamma_{\ep N}(F^{l}\bk_0)$, considering two cases for 
$0\neq\bk_0\in\IZ^{2d}_{N}$:

\begin{itemize}
\item Assume that $F^{l}\bk_{0}\in \IZ^{2d}_{N}$ for all $l=0,\ldots,n-1$. 
From Proposition \ref{thmart} and the assumptions on $n$, we have
\bea
\label{cm}
\sum_{l=0}^{n-1}|F^{l}\bk_{0}|^{2} \geq
\min_{0 \not = \bk \in \IZ^{2d}} \sum_{l=0}^{n-1}|F^{l}\bk|^{2}>
e^{2(1-\del')\hat{h}(n-1)}> n\, e^{2(1-\del)\hat{h}(n-1)}.
\eea
Thus for any such $\bk_{0}$, there exists $l_0\in\{0,\ldots,n-1\}$ such that
$|F^{l_0}\bk_{0}|>e^{(1-\del)\hat{h}(n-1)}$.

\item Assume there exists $0\leq l_0\leq n-1$ 
such that $\{\bk_0,\ldots,F^{l_0}\bk_0\}\in\IZ^{2d}_N$, while 
$F^{l_0+1}\bk_0\not\in\IZ^{2d}_N$. 
Using the RHS of (\ref{Et1}), we necessarily have
$|F^{l_0}\bk_0|\geq \frac{N}{2\, \|F\|}>e^{(1-\del)\hat{h}(n-1)}$.
\end{itemize}
Gluing together both cases and using \eqref{pGauss}, we infer that for any 
$0\neq \bk_{0}\in \IZ^{2d}_{N}$, there is an index $0\leq l_0\leq n-1$ such that
\bean
\gamma_{\ep,N}(F^{l_0}\bk)
\leq \exp\big\{-\ep^2\, e^{2(1-\del)\hat{h}(n-1)}\big\}+ Ce^{-\frac{(\ep N)^2}{4}}
< e^{-4}+Ce^{-\frac{M^2}{4}}.
\eean
From the first condition in \eqref{M}, the RHS is $<e^{-1}$, so that $n\geq\tau_q(\ep,N)$.
This holds for any $n$ satisfying \eqref{Et1}. We have proven that in the r\'egime
$\ep N>M$, one has
$\tau_q(\ep,N)\leq \frac{\ln (2\,\ep^{-1})}{(1-\del)\hat{h}}+2$.
This is true for any $\del>0$ and sufficiently small $\ep$, which ends the proof of {\it II)}.
\qed


\medskip

{\bf Acknowledgments:} The third author would like to thank Prof. B. Nachtergaele
for helpful suggestions and discussions.  We thank the referee for suggesting the
comparison between the relaxation time scale and the decay of quantum fidelity.

\appendix

\section{Proofs of some elementary facts}


\subsection{Proof of Proposition \ref{qnoise}}
\label{pqnoise}
The value of the normalization constant is computed as follows
\bean
Z=\sum_{\bn\in \IZ^{2d}_{N}} \tilde{g}_{\ep}(N^{-1}\bn)=
\sum_{\bn\in \IZ^{2d}} g_{\ep}(N^{-1}\bn)
=N^{2d}\sum_{\bn\in\IZ^{2d}}g_{\ep N}(\bn)= N^{2d}\tilde{g}_{\ep N}(0).
\eean
Using the periodicity $ad(W_{\bn+N\bm})=ad(W_{\bn})$, the quantum noise operator can
be expressed as:
\bean
\cG_{\ep,N}&=& 
\frac{1}{Z}\sum_{\bn\in \IZ^{2d}_{N}}\tilde{g}_{\ep} \lp\frac{\bn}{N}\rp ad(W_{\bn})=
\frac{1}{N^{2d}\tilde{g}_{\ep N}(0)}\sum_{\bn\in \IZ^{2d}}g_{\ep} \lp\frac{\bn}{N}\rp ad(W_{\bn})\\
&=&\frac{1}{\tilde{g}_{\ep N}(0)}\sum_{\bn\in\IZ^{2d}}g_{\ep N}(\bn)\,ad(W_{\bn}).
\eean
Applying the commutation relations \eqref{MCCR}, $\cG_{\ep,N}$ acts on $W_{\bk}$ as follows
\bean
\cG_{\ep,N}W_{\bk}
=\frac{1}{\tilde{g}_{\ep N}(0)}\sum_{\bn\in\IZ^{2d}}g_{\ep N}(\bn)\,
e^{\frac{2\pi i}{N} \bk\wedge \bn}\,W_{\bk}. 
\eean 
\qed

\subsection{Proof of Lemma \ref{l:chainrule}\label{a:chainrule}}
The first assertion can be proven along the lines of \cite[Lemma 2.2]{BouzRob}, by 
an induction argument over the degree $k$ of differentiation (the only difference is that
our map is defined for discrete times). 

Our induction hypothesis: for any $0\leq k'<k$ there exists $\tilde{C}_{k'}$ such that 
for any multiindex $|\gamma|=k'$,
$|\partial^\gamma\Phi^t|\leq \tilde{C}_{k'}\,e^{\Gamma k't}$. 
The case $k=1$ is obvious: $\|\Phi^t(\bx)\|\leq C$ uniformly in time.

We now take a multiindex $\al,\ |\al|=k$, and apply the chain rule:
$$
\partial^\al (\Phi\circ\Phi^{t})= 
\sum_{j=1}^{2d} (\partial_j\Phi)\circ\Phi^t\times \partial^\al(\Phi^t)_j
+\sum_{\gamma\leq\al,|\gamma|>1}(\partial^\gamma\Phi)\circ\Phi^t\times\cB_{\al,\gamma}(\phi^t).
$$
Here $\cB_{\al,\gamma}(\phi^t)$ is a sum of products of derivatives of $\Phi^t$ of order 
$<k$; using the induction hypothesis, each product is $\leq C\,e^{\Gamma kt}$.
Now we use the discrete-time version of \cite[Lemma 2.3]{BouzRob}. Namely,
for a given point $\bx$, the above equation may be written
$$
X(t+1)=M(t)X(t)+Y(t),
$$
where $X(t)=\partial^\al(\Phi^t)(\bx)$ is ``unknown'', 
the matrix $M(t)=D\Phi(\Phi^t(\bx))$ satisfies
$\|M(t)\|\leq e^\Gamma$ for all times, and we checked above that $\|Y(t)\|\leq C\,e^{\Gamma kt}$. 
From the explicit expression
$$
X(t+1)=\big(\prod_{s=1}^t M(s)\big)X(1) + \big(\prod_{s=2}^t M(s)\big)Y(1)
+ \big(\prod_{s=3}^t M(s)\big)Y(2)+\ldots +Y(t),
$$
one easily checks that $\|X(t)\|\leq \tilde{C_k}\,e^{\Gamma kt}$ for a certain constant $\tilde{C_k}$,
which proves the induction at the order $k$.
Composing $\Phi^t$ with an observable $f$, we easily get the first assertion of the Lemma.

To get the second assertion, we notice that the noise operator consists in
averaging over maps of the type 
$\Phi^t_{\{\bv_j\}}=t_{\bv_t}\Phi t_{\bv_{t-1}}\Phi\cdots t_{\bv_1}\Phi$. Now, one can easily
adapt the above proof to show that each of those
maps satisfies, for $|\al|=k$,
$$
|\partial^\al(f\circ\Phi_{\{\bv_j\}}^{t})|\leq \tilde{C}_k\,\|f\|_{C^k}\,e^{\Gamma kt},
$$ 
with $\tilde{C}_k$ independent
of the realization $\{\bv_j\}$. Averaging over the realizations does not harm the upper bound, 
yielding the second assertion.\qed


\subsection{Proof of Lemma \ref{GES}}
\label{est}
Since the $2d$-dimensional Gaussian $e^{-|\bxi|^2}$ factorizes into $\prod_i e^{-\xi_i^2}$,
it is natural to first treat the one-dimensional case, that is consider the periodized
Gaussian (a Jacobi theta function)
$$
\theta_\sigma(\xi)
=\sum_{\nu\in\IZ} e^{-\sigma^2(\xi+\nu)^2},\qquad 
\tilde\theta_\sigma(\xi)=\sum_{0\neq\nu\in\IZ} e^{-\sigma^2(\xi+\nu)^2}.
$$
If we assume that $\xi\in(-1/2,1/2]$, one has $\nu+{\xi}>\nu-1/2$ for $\nu>0$ and
$\nu+{\xi}<\nu+1/2$ for $\nu<0$. From the monotonicity of the Gaussian on
$\IR^{\pm}$, this implies
\bea\label{e:1dJacobi}
\tilde\theta_\sigma(\xi)\leq\theta_\sigma(1/2)
=2\,e^{-\sigma^2/4}\sum_{\nu\geq 0}e^{-\sigma^2\nu(\nu+1)}
\leq 2\,e^{-\sigma^2/4}\,\theta_\sigma(0).
\eea
We will also use the lower bound:
\bea\label{e:jacobi}
\theta_\sigma(\xi)=e^{-\sigma^2\xi^2}
\big(1+\sum_{\nu>0} 2\cosh(2\sigma^2\nu\xi)\,e^{-\sigma^2 \nu^2}\big)
\geq e^{-\sigma^2\xi^2}\,\theta_\sigma(0).
\eea
We can now pass the the $2d$-dimensional case and consider $\bxi$, with all
components in $(-1/2,1/2]$. An easy bookkeeping shows that
\begin{multline}
\theta_\sigma({\bxi})
:=\prod_{i=1}^{2d}\theta_\sigma({\xi}_i)=e^{-\sigma^2|{\bxi}|^2}
+\tilde\theta_\sigma({\xi}_1)\prod_{i=2}^{2d}\theta_\sigma({\xi}_i)
+e^{-\sigma^2{\bxi}_1^2}\tilde\theta_\sigma({\xi}_2)
\prod_{i=3}^{2d}\theta_\sigma({\xi}_i)\\
+e^{-\sigma^2({\bxi}_1^2+{\bxi}_2^2)}\tilde\theta_\sigma({\xi}_3)
\prod_{i=4}^{2d}\theta_\sigma({\xi}_i)
+\ldots
+e^{-\sigma^2({\bxi}_1^2+\ldots+{\bxi}_{2d-1}^2)}\tilde\theta_\sigma({\xi}_{2d}).
\end{multline}
Using the bound (\ref{e:1dJacobi}) and the fact that the maximum of $\theta_\sigma$
is $\theta_\sigma(0)>1$, we obtain:
\bea\label{e:jacobi2}
\theta_\sigma({\bxi})\leq e^{-\sigma^2|{\bxi}|^2}+
4d\,e^{-\sigma^2/4}\,\theta_\sigma(0)^{2d}
=e^{-\sigma^2|{\bxi}|^2}+4d\,e^{-\sigma^2/4}\,\theta_\sigma({\bf 0}).
\eea
The quantum eigenvalues are expressed in terms of the function 
$$
\gamma_{\ep,N}(\bxi)=\gamma_{\ep,N}(\bxi^N)
=\frac{\theta_{\ep N}(\bxi^N/N)}{\theta_{\ep N}({\bf 0})}.$$
From the estimates (\ref{e:jacobi},\ref{e:jacobi2}), this function satisfies
$$
e^{-\ep^2|\bxi^N|^2}\leq\gamma_{\ep,N}(\bxi^N)
\leq \frac{e^{-\ep^2|{\bxi}^N|^2}}{\theta_{\ep N}({\bf 0})}+4d\,e^{-(\ep N)^2/4}
\leq e^{-\ep^2|{\bxi}^N|^2}+4d\,e^{-(\ep N)^2/4}.
$$
\qed
\section{Egorov estimates}

\subsection{Proof of Proposition \ref{p:Egorov}\label{a:Egorov}}
We need to prove the statement for one iterate of the map ($n=1$).
As explained in Section~\ref{q-maps}, $\Phi$ is a combination of a 
linear automorphism $F$, a translation $t_{\bv}$ and
the time-1 flow map $\Phi_1$: $\Phi=F\circ t_{\bv} \circ \Phi_{1}$. The 
quantum propagator on $\cA_N$ is given by the (contravariant) product:
\begin{equation}\label{fullmap}
\cU(\Phi)=\cU(\Phi_1)\cU(t_{\bv})\cU(F).
\end{equation}
We estimate the quantum-classical discrepancy
of each component separately. The estimate will be valid for either the operator
norm on $\cH_{N,\btht}$, or the Hilbert-Schmidt norm.

As explained in Section~\ref{catmap}, the correspondence is exact for the
linear automorphism: 
\begin{equation}\label{Ego1}
\cU(F)Op(f)=Op(f\circ F).
\end{equation}

The translation $t_{\bv}$ is quantized by a quantum translation of vector $\bv^{(N)}$,
which is at a distance $|\bv-\bv^{(N)}|\leq  CN^{-1}$: $\cU(\bv)Op(f)=Op(f\circ t_{\bv^{(N)}})$. 
If we Fourier decompose $f=\sum_{\bk}\hat f(\bk)w_{\bk}$, we have trivially
$f\circ t_{\bv}=\sum_{\bk} e^{2i\pi \bk\wedge\bv}\,f(\bk)\,w_{\bk}$. As a result, 
since for our norms $\|W_{\bk}\|=1$, we simply get
$$
\|\cU(\bv)Op(f)-Op(f\circ t_{\bv})\|\leq\sum_{\bk}|f(\bk)|\,|e^{2i\pi \bk\wedge\bv^{(N)}}
-e^{2i\pi \bk\wedge\bv}|.
$$
The last factor in the RHS is an $\cO\big(\frac{|\bk|}{N}\big)$. 
Since the Fourier coefficients decay as $|\hat{f}(\bk)|\leq C_M\frac{\|f\|_{C^M}}{(1+|\bk|)^M}$
for any $M>0$, we can take $M=2d+2$, which makes the sum over $\bk$ finite,
and we obtain
\begin{equation}\label{Ego2}
\|\cU_N(\bv)Op_N(f)-Op_N(f\circ t_{\bv})\|\leq C\frac{\|f\|_{C^{2d+2}}}{N}.
\end{equation}

The quantum-classical discrepancy due to the nonlinear map $\Phi_1$ is computed along the
lines of \cite{BouzRob}. $\Phi_1$ is time-1 map generated by the flow of Hamiltonian $H(t)$. We want to
compare $Op(f\circ\Phi_1)$ with the quantum-mechanically evolved observable
$\cU(\Phi_1) Op(f)$. To do so, we compare the infinitesimal evolutions.
Let us call 
$\cU(t,s)=ad\big(\cT\,e^{-\frac{i}{\hbar}\int_s^t Op(H(r))dr}\big)$ the quantum 
propagator between times $s<t$, and $K(t,s)$ the corresponding classical propagator.
Duhamel's principle lies in the following observation: from the identities
$$
\frac{d}{dt}\cU(t,s)A=i\hbar^{-1}\cU(t,s)[Op(H(t)),A],\qquad 
\frac{d}{ds}K(t,s)f=-\{H(s),K(t,s)f\},
$$
one constructs the following total derivative:
\begin{equation}\label{integrand}
\frac{d}{dt}\big(\cU(t,0) Op(K(1,t)f)\big)
=\cU(t,0)\left\{i\hbar^{-1}\big[Op(H(t)),Op(K(1,t)f)\big]-Op\big(\{H(t),K(1,t)f\}\big)\right\}.
\end{equation}
Integrating over $t\in[0,1]$ and taking the norm, using the unitarity of $\cU(t,0)$, one gets:
\begin{equation}
\|\cU(\Phi_1) Op(f)-Op(K(1,0)f)\|\leq
\int_0^1 dt\,\Vert i\hbar^{-1}[Op(H(t)),Op(K(1,t)f)]-Op\big(\{H(t),K(1,t)f\}\big)\Vert.
\end{equation}
We can easily estimate the norm of \eqref{integrand}, using the Fourier decomposition
of $H(t)$ and $K(1,t)f$: 
we write $H(t)=\sum_{\bk}\hat{H}(\bk,t)\,w_{\bk}$, 
$K(1,t)f=\sum_{\bm}\hat f(\bm,t)\,w_{\bm}$, and expand. The CCR \eqref{MCCR}
and their corresponding Poisson brackets read
$$
[W_{\bk},W_{\bm}]=2i\sin(\pi \bk\wedge\bm/N)\,W_{\bk+\bm},\quad 
\{w_{\bk},w_{\bm}\}=-4\pi^2 \bk\wedge\bm\,w_{\bk+\bm}.
$$
This gives us for the operator in the above integral:
$$
\sum_{\bk,\bm}\hat{H}(\bk,t)\,\hat f(\bm,t)\,4\pi\big\{\pi\bk\wedge\bm-N\sin(\pi \bk\wedge\bm/N)\big\}
W_{\bk+\bm}.
$$
The term in the curly brackets is an $\cO\big(\frac{(|\bk||\bm|)^2}{N}\big)$, while
the product of Fourier coefficients decays like $(|\bk|\,|\bm|)^{-M}$ for any
$M>0$, due to the smoothness of $H(t)$ and $f$. 
To be able to sum over $\bk,\ \bm$ we need to take $M\geq 2d+3$, and get for 
any $t\in[0,1]$:
\bea
\| i\hbar^{-1}[Op(H(t)),Op(K(1,t)f)]-Op\big(\{H(t),K(1,t)f\}\big)\|\leq 
C\frac{\|H(t)\|_{C^M} \|K(1,t)f\|_{C^M}}{N}.
\eea 
Due to the smoothness of $H(t)$, the norm $\|K(1,t)f\|_{C^M}$ can only differ from $\|f\|_{C^M}$
by a finite factor independent of $f$ \cite{BouzRob}.
We therefore get for any smooth $f$:
\begin{equation}\label{Ego3}
\|\cU(\Phi_1) Op(f)-Op(f\circ\Phi_1)\|\leq C\frac{\|f\|_{C^{2d+3}}}{N}.
\end{equation}

We now control the quantum-classical discrepancy stepwise. We 
use the discrete-time Duhamel principle to control the discrepancy for the full map \eqref{fullmap}:
\begin{multline}
\|\cU(\Phi)Op(f)-Op(f\circ\Phi)\|\leq \|\cU(F)Op(f)-Op(f\circ F)\|+\\ 
+\|\cU(t_{\bv})Op(f\circ F)-Op(f\circ F\circ t_{\bv})\|+
\|\cU(\Phi_1)Op(f\circ F\circ t_{\bv})-Op(f\circ F\circ t_{\bv}\circ \Phi_1)\|,
\end{multline}
and for its iterates:
\bea
\|\cU(\Phi)^nOp(f)-Op(f\circ\Phi^n)\|
\leq \sum_{j=0}^{n-1} \|\cU(\Phi)Op(f\circ\Phi^j)-Op(f\circ\Phi^{j+1})\|.
\eea
Putting together the estimates (\ref{Ego1},\ref{Ego2},\ref{Ego3}) we get the 
statement of the Proposition, with either norm $\|\cdot\|_{\cB(\cH_N)}$ or $\|\cdot\|_{HS}$.
\qed


\subsection{Proof of Proposition \ref{p:Egorov-noisy}\label{a:Egorov-noisy}}
Compared with the previous appendix, we now also need to control the discrepancy between classical
and quantum noise operators. This is quite easy to do in Fourier space: for any
$f\in C^\infty(\IT^{2d})$, we have:
\begin{equation}\label{e:noise-corresp}
\|\cG_{\ep,N}Op(f)-Op(G_\ep f)\|\leq 
\sum_{\bk\in\IZ^{2d}}|\gamma_{\ep, N}(\bk)-\hat g_\ep(\bk)|\,|\hat f(\bk)|.
\end{equation}
 Let us assume that the Fourier transform of $g$ decays as
$|\hat g(\bxi)|\leq \frac{C}{(1+|\bxi|)^{D}}$ as $\bxi\to\infty$, with $D\geq 2d+1$. 
From the explicit expression \eqref{PSFgamma}, we easily get
the estimate (in the limit $\ep N\to\infty$):
\bea\label{e:est-gamma}
\gamma_{\ep, N}(\bk)=
\frac{\hat g_\ep(\bk)+\sum_{\bm\neq 0}\cO\big((\ep N|\bm|)^{-D}\big)}
{\hat g_\ep(0)+\sum_{\bm\neq 0}\cO\big((\ep N|\bm|)^{-D}\big)}
=\hat g_\ep(\bk)+\cO\big((\ep N)^{-D}\big),
\eea
and the estimate is uniform for $\bk\in \IZ_N^{2d}$. For $\bk$ outside $\IZ_N^{2d}$, we simply
bound the difference by 
$$
|\gamma_{\ep, N}(\bk)-\hat g_\ep(\bk)|\leq 2.
$$
Therefore, for any $f\in C^\infty_0(\IT^{2d})$, one has:
\begin{equation}
\|\cG_{\ep,N}Op(f)-Op(G_\ep f)\|\leq 
\sum_{\bk\in\IZ_N^{2d}-0}\frac{C}{(\ep N)^D}\,|\hat f(\bk)|+
2\sum_{\bk\in\IZ^{2d}\setminus\IZ_N^{2d}}|\hat f(\bk)|\leq C\frac{\|f\|_{C^D}}{(\ep N)^D}.
\end{equation}

From the previous appendix we control the quantum-classical discrepancy of the
unitary step $\cU(\Phi)$. Both yield:
\begin{align*}
\|\cT_{\ep,N}Op(f)-Op(T_\ep f)\|&\leq \|\cG_{\ep,N}\big(\cU(\Phi)Op(f)- Op(K_\Phi f)\big)\|
+\|\cG_{\ep,N}Op(K_\Phi f)-Op(G_\ep K_\Phi f)\|\\
&\leq \|\cU(\Phi)Op(f)- Op(K_\Phi f)\|+\|\cG_{\ep,N}Op(K_\Phi f)-Op(G_\ep K_\Phi f)\|\\
&\leq C\frac{\|f\|_{C^{2d+3}}}{N}+C\frac{\|f\|_{C^D}}{(\ep N)^D},
\end{align*}
valid for any $D\geq 2d+1$.
To obtain the Proposition, we apply an obvious generalization of Duhamel's principle, 
using the fact that $\cT_{\ep,N}$ is contracting on
$\cA_N^0$.\qed



\begin{thebibliography}{99}

\bibitem{AF94} R.~Alicki  and M.~Fannes, 
{\em Defining quantum dynamical entropy} {\sl Lett.~Math.~Phys.} {\sf 32} (1994) 75-82

\bibitem{ALPZ}
R.~Alicki,  A.~{\L}ozi{\'n}ski,
P.~Pako{\'n}ski and K.~{\.Z}yczkowski:
{\em Quantum dynamical entropy and decoherence rate}.
{\em J. Phys} {\sf A 37} (2004) 5157-5172

\bibitem{Arnold}
V.I.~Arnold  and A.~Avez: {\em Ergodic Problems of Classical Mechanics}.
The Mathematical Physics Monograph Series, W.A. Benjamin, 1968.

\bibitem{B}
V.~Baladi: {\em Positive Transfer Operators and Decay of Correlations}.
Advanced Series in Nonlinear Dynamics vol. 16 , World Scientific, 2000.

\bibitem{BCCFV} 
F.~Benatti , V.~Cappellini, M.~De Cock, M.~Fannes and
D.~Vanpeteghem, {\em Classical Limit of Quantum Dynamical Entropies.}
{\sl Rev. Math. Phys.} {\sf 15} (2003), no. 8, 847--875.

\bibitem{BPS}  
P.~Bianucci , J.~P.~Paz, and  M.~Saraceno,
{\em Decoherence for classically chaotic quantum maps}.
{\sl Phys.~Rev.} {\sf E 65}, 046226 (2002).

\bibitem{BKL}
M.~Blank, G.~Keller and C.~Liverani: {\em Ruelle-Perron-Frobenius spectrum for Anosov maps}.
Nonlinearity {\sf 15} (2002) 1905-1973

\bibitem{BonDB1}
F.~Bonechi  and S.~De~Bi\`{e}vre: {\em Exponential mixing and $ln \hbar$ timescales in
quantized hyperbolic maps on the torus}. {\em Commun. Math. Phys.} {\sf 211} (2000) 659--686

\bibitem{BonDB2}
F.~Bonechi and S.~De~Bi\`{e}vre: {\em Controlling strong scarring for quantized ergodic
toral automorphisms}. {\em Duke Math. J.}  {\em 117}  (2003) 571--587

\bibitem{BozDB}
A.~Bouzouina  and S.~De~Bi\`{e}vre: {\em Equipartition of the eigenfunctions of quantized
ergodic maps on the torus}. {\em Commun. Math. Phys.} {\sf 178} (1996) 83--105

\bibitem{BouzRob}
A.~Bouzouina  and D.~Robert: {\em Uniform Semi-classical Estimates for the Propagation of
Quantum Observables}. {\em Duke Math. J.} {\sf 111} (2002) 223--252 

\bibitem{Br}
D.~Braun: {\em Dissipative Quantum Chaos and
Decoherence}, Springer Tracts in
 Modern Physics {\sf 172}, Springer, Heidelberg (2001)

\bibitem{caldeira-leggett}A.O. Caldeira and A.J. Leggett: {\em Influence of damping on quantum 
interference: An exactly soluble model}. {\em Phys. Rev.} {\sf A 31} (1985) 1059--1066

\bibitem{CC}
G.~Casati and B.~Chirikov: {\em Quantum Chaos. Between Order and Disorder.}
 Cambridge University Press, Cambridge (1999).

\bibitem{cerruti}
N.R.~Cerruti and S.~Tomsovic: {\em A uniform approximation for the fidelity in chaotic
systems}, J. Phys. {\sf A 36} (2003) 3451--3465; 
Corrigendum, J. Phys. {\sf A 36} (2003) 11915--11916.

\bibitem{CZ}
C.~Conley and E.~Zehnder: {\em The Birkhoff-Lewis fixed point theorem and a conjecture of 
V.I. Arnold}, {\em Invent. Math.} {\sf 73} (1983) 33--49

\bibitem{davies}
E.B.~Davies, {\em Semigroup growth bounds}, to appear in {\em J.Oper.Theory},
arXiv:math.SP/0302144 (2003)

\bibitem{DB}
S.~De~Bi\`{e}vre: {\em Chaos, quantization and the classical limit on the torus}.
Proceedings of the XIVth Workshop on Geometrical Methods in Physics. Bia{\l}owie\.{z}a 1995,
mp$\_$arc 96-191, PWN (Polish Scientific Publisher) 1998.

\bibitem{DE}
M.~Degli~Esposti: {\em Quantization of the orientation preserving automorphisms of the torus}.
{\em Ann. Inst. Henri Poincar\'{e}} {\sf 58} (1993) 323-341.

\bibitem{DEGI}
M.~Degli~Esposti, S.~Graffi and S.~Isola: {\em  Classical limit
of the quantized hyperbolic toral automorphisms}, {\em Commun. Math. Phys.} {\sf 167} (1995)  471--507

\bibitem{F2}
A.~Fannjiang: {\sl Time Scales in Noisy Conservative
Systems}.
Lecture Notes in Physics {\sf 450} (1995) 124-139

\bibitem{F}
A.~Fannjiang: {\sl Time scales homogenization of Periodic Flows with Vanishing
Molecular Diffusion}.
{\em Jour.  Diff. Eq.} {\sf 179} (2002) 433-455

\bibitem{FNW}
A.~Fannjiang, S.~Nonnenmacher and L.~Wo{\l}owski: {\em Dissipation time and
decay of correlations}. {\em Nonlinearity} {\sf 17} (2004) 1481-1508

\bibitem{FW}
A.~Fannjiang and L.~Wo{\l}owski: {\em Noise Induced Dissipation in Lebesgue-Measure
Preserving Maps on d-Dimensional Torus}. {\em Journal of Statistical Physics}
Vol. {\sf 113} (2003) 335-378

\bibitem{FNDB}
F.~Faure, S.~Nonnenmacher and S.~De~Bi\`evre: {\em Scarred eigenstates for quantum cat maps
of minimal periods}. {\em Commun. Math. Phys.} {\sf 239} (2003) 449--492

\bibitem{fishman}
S. Fishman and S. Rahav, {\em Relaxation and Noise in Chaotic Systems}, Lecture notes, 
Ladek winter school (2002), nlin.CD/0204068

\bibitem{GS}
I.~Garcia-Mata and M.~Saraceno:  
{\em Spectral properties and classical decays in quantum open systems}, 
{\em Phys. Rev.} {\sf E 69} (2004) 056211

\bibitem{GSS}
I.~Garcia-Mata, M.~Saraceno and M.-E.~Spina: 
\emph{Classical decays in decoherent quantum maps.} {\em Phys. Rev. Lett.} {\sf 91} (2003) 064101

\bibitem{gardiner}
C.W.~Gardiner, {\em Quantum noise}. {\em Springer Series in Synergetics} {\sf 56}, 
Springer, Heidelberg (1991)

\bibitem{GL}
S.~Gou\"ezel and C.~Liverani: {\em Banach spaces adapted to Anosov systems}. Preprint 
arXiv:math.DS/0405278

\bibitem{GraDE}
S.~Graffi and M.~Degli Esposti (Eds.): {\em The Mathematical Aspects of Quantum Maps}.
{\em Lecture Notes in Physics} {\sf 618}, Springer, Heidelberg (2003)

\bibitem{Haake}
F.~Haake: {\em Quantum Signatures of Chaos}, Springer, Heidelberg (2000)

\bibitem{HB}
J.H.~Hannay and M.V.~Berry: {\em Quantization of linear maps on a torus
- Fresnel diffraction by a periodic grating}.
{\em Physica D} {\sf 1} (1980) 267-290

\bibitem{jalabert}
R.A.~Jalabert and H.M.~Pastawski, {\em Environment-Independent Decoherence Rate in
Classically Chaotic Systems}, Phys. Rev. Lett. {\sf 86} (2001) 2490--2493

\bibitem{K}
J.P.~Keating: {\em  The cat maps: quantum mechanics and classical motion}.
{\em Nonlinearity}  {\sf 4} (1991) 309-341

\bibitem{KMR}
J.P.~Keating, F.~Mezzadri and  J.M.~Robbins: {\em Quantum boundary conditions
for torus maps}.
{\em Nonlinearity}  {\sf 12} (1991) 579-591

\bibitem{Ki}
Yu.~Kifer: {\em Random Perturbations of Dynamical Systems}.
Birkh\"{a}user, Boston, 1988.

\bibitem{lindblad} G. Lindblad, {\em On the generators of quantum dynamical semigroups}, 
{\em Commun. Math. Phys.} {\sf 48} (1976) 119--130;
K.~Kraus: {\em States, Effects and Operations: Fundamental Notions of Quantum Theory}.
Springer, Berlin, 1983

\bibitem{Liv}C.~Liverani, {\em On contact Anosov flows}, to appear in {\em Annals of 
Mathematics}, Preprint math.DS/0303237

\bibitem{manderfeld}C.~Manderfeld, J.~Weber and F.~Haake: 
\emph{Classical versus quantum time evolution of (quasi-) probability densities at 
limited phase-space resolution.} {\em J. Phys.} {\sf A 34} (2001) 9893--9905 

\bibitem{MR}
J.~Marklof and Z.~Rudnick:{\em Quantum unique ergodicity for parabolic maps},
{\em Geom. Funct. Anal.} {\sf 10} (2000) 1554-1578

\bibitem{M}
F.~Mezzadri: {\em On the multiplicativity of quantum cat maps}.
{\em Nonlinearity} {\sf 15} (2002) 905-922

\bibitem{N}
S.~Nonnenmacher: {\em Spectral properties of noisy classical and quantum propagators.}
{\em Nonlinearity} {\sf 16} (2003) 1685-1713


\bibitem{PZ}
J.-P.~Paz and W.H.~Zurek: {\em Decoherence, Chaos and the Second Law}. {\em Phys. Rev. Lett.}
{\sf 72} (1994) 2508-2511

\bibitem{peres}
A.~Peres, {\em Stability of quantum motion in chaotic and regular systems},
Phys. Rev. {\sf A 30} (1984) 1610--1615

\bibitem{prosen}
T.~Prosen and M.~{\v Z}Znidari{\v c}, {\em Stability of quantum motion and correlation decay},
J. Phys. {\sf A 35} (2002) 1455--1481

\bibitem{RSO}
A.M.F.~Rivas, M.~Saraceno and A.M.~Ozorio de Almeida: {\em Quantization of 
multidimensional cat maps}. {\em Nonlinearity} {\sf 13} (2000) 341--376

\bibitem{Ro}
D.~Robert: {\em Autour de l'approximation semi-classique}. Birkh\"auser, Boston, 1987.

\bibitem{beenakker}
P.G.~Silvestrov, J.~Tworzyd{\l}o and C.W.J.~Beenakker, {\em Hypersensitivity to
perturbations of quantum-chaotic wave-packet dynamics}, Phys. Rev. {\sf E 67} (2003)
025204


\bibitem{Zas}
G.M.~Zaslavsky: 
{\em Stochasticity in quantum systems}. {\em Phys. Rep.} {\sf 81}
(1981) 157--250

\bibitem{Z}
S.~Zelditch:
{\em Index and dynamics of quantized contact transformations}. 
{\em Annales de l'institut Fourier}, {\sf 47} no. 1 (1997)  305--363

\end{thebibliography}
\end{document}